\documentclass[twocolumn,showpacs]{revtex4}
\usepackage{graphicx,epsf,bm,amsmath,color}
\def\bk{\mbox{\boldmath $k$}}

\def\bq{\mbox{\boldmath $q$}}
\def\br{\mbox{\boldmath $r$}}

\def\bfsigma{\mbox{\boldmath $\sigma$}}
\def\bftau{\mbox{\boldmath $\tau$}}
\newcommand{\ltsim}{\protect\raisebox{-0.5ex}{$\:\stackrel{\textstyle <}
	{\sim}\:$}}

\begin{document}
\title{Nuclear and neutron matter $G$-matrix calculations with Ch-EFT potential
including effects of three-nucleon interaction}
\author{M. Kohno\footnote{kohno@kyu-dent.ac.jp}}
\affiliation{Physics Division, Kyushu Dental University,
Kitakyushu 803-8580, Japan}

\begin{abstract}
Energies of symmetric nuclear matter and neutron matter are evaluated
in the lowest order Bruekner theory using the Ch-EFT potential including
effects of the three-nucleon force (3NF). The 3NF is first reduced to
density-dependent nucleon-nucleon (NN) force by folding single-nucleon
degrees of freedom in infinite matter. Adding the reduced NN force to the
initial NN force and applying a partial-wave expansion, we perform $G$-matrix
calculations in pure neutron matter as well as in symmetric nuclear.
We obtain the saturation curve which is close to the empirical one.
It is explicitly shown that the cutoff-energy dependence of the calculated
energies is substantially reduced by including the 3NF. Characters of the
3NF contributions in separate spin and isospin channels are discussed.
Calculated energies of the neutron matter are very similar to those
used in the literature for considering neutron star properties.
\end{abstract}
\pacs{21.30.Fe, 21.65.Cd, 21.65.Mn, 26.60.Kp}

\maketitle

\section{Introduction}
One of the basic problems in nuclear physics is to understand characteristic
properties of nuclei, especially the saturation and single-particle
shell structure, starting from underlying nucleon-nucleon (NN) interactions. 
Various many-body theories have been developed since the 1950s.
The Brueckner theory \cite{BLM54, DAY67, BET71}, which was initiated in
a multiple-scattering viewpoint and later established as the linked-cluster
expansion in terms of $G$-matrices, has served as an standard
method to understand nuclei as the system of nucleons moving independently
in a mean field in spite of the NN interactions having singularly strong repulsion
in the short-ranged region. Another basic method for quantum many-body
problems is a variational treatment \cite{PW79}, although shell structure is not
intuitive in this framework. Two methods in a non-relativistic framework are
now known to provide similar description of nuclear bulk properties \cite{BB12}.
These results unfortunately indicate that the saturation cannot
be correctly reproduced in a non-relativistic framework when realistic
NN potentials are used. Various elaborate many-body methods
practiced in recent years, such as the coupled-cluster method, the unitary
correlated method, and the no-core shell model with low-momentum interactions,
confirm this situation.

Many attempts have been made to find other mechanisms to improve the
description of the saturation properties, such as relativistic effects and
three-nucleon force (3NF) contributions in the nuclear medium. It was
demonstrated that relativistic Brueckner-Hartree-Fock calculations \cite{BM90}
can provide a satisfactory saturation curve. However, because contributions
from higher-order correlations and three-nucleon forces have not been fully
estimated in the relativistic treatment, the problem seems not to be settled yet.
In the last decade, a new description of the NN potential has been
developed \cite{MACH,EGM05,EHM09};
that is, the interaction based on chiral effective field theory (Ch-EFT).
The potential form is dictated by underlying chiral symmetry of the QCD, and
potential parameters, low-energy constants, are adjusted to explain
scattering data as in other realistic NN potentials. 3NFs
are systematically introduced in this framework and most parameters
for these terms are taken over from the NN sector.

The introduction of 3NFs has a long history since the 1950s, and various
studies have been devoted for 3NF contributions in nuclear properties.
Besides three-body correlations through ordinary NN forces, 3NFs may
arise from excitations of virtual nucleon-antinucleon pairs as well as
isobar $\Delta$ and other nucleon excited states in the medium
\cite{KL53,FM57}. The importance of 3NFs, in the standard non-relativistic
description of nuclei, has been established by precise few-body
calculations \cite{WIR00,KNE12}. It has also been recognized that 3NFs
are necessary to reproduce empirical saturation properties
\cite{BB12,KAT74,FP81,APR98}. Although numerical calculations of
reproducing nuclear saturation properties by including 3NF effects
have been presented by many authors, the advantage of using the the
Ch-EFT is that the contribution of the 3NF can be discussed in a way
systematic and consistent with the initial NN interaction. 

Some perturbative considerations for neutron matter properties with
the Ch-EFT interaction including the 3NF have been reported
in Refs. \cite{BSFN05,HS10,HEB11,TKHS13,CORA13}.
The present author gave, in Ref. \cite{MK12} a brief report of the
lowest-order Brueckner theory (LOBT) calculation in nuclear matter
with using the reduced density-dependent NN force
obtained from the Ch-EFT 3NF, in which
a focus was put on the effective spin-orbit strength. 
Similar LOBT calculations also appeared in Ref. \cite{SCC12}. 

In this paper, we report, in details, nuclear and neutron matter calculations
in the LOBT based on the N$^3$LO Ch-EFT
potential including its N$^2$LO 3NF. Because the Ch-EFT is a definite way to
organize the interaction between nucleons, it is important to study the
implication of the interaction based on it to the nuclear many-body problem.
However, it is currently impossible to consider full contributions
of 3NFs together with many-body correlations,
except for very light nuclei.
Even for the NN force, it is already very difficult to take into
account effects of more than three-nucleon correlations.
Therefore, we introduce an approximation. First, reduced
effective NN forces are constructed by averaging the 3NF
over the third nucleon in the Fermi sea. Adding the reduced NN force
to the initial Ch-EFT NN interaction, we carry out standard $G$-matrix
calculations. This procedure may not be accommodated rigorously in a
linked-cluster expansion of the quantum many-body theory. Nevertheless,
we should expect meaningful information about the role of 3NFs
in this estimation.

The procedure of LOBT calculations with including the reduced
NN force from the Ch-EHF 3NF is explained in Sect. II.
For numerical calculations, it is necessary to make a partial-wave expansion
of the reduced NN interaction. This is straightforward but somewhat intricate.
Explicit expressions of the reduced NN interaction are given in Appendix A.
Expressions after the partial-wave expansion are
shown in Appendix B. Numerical results are presented first for nuclear matter in Sec. III, 
and then for neutron matter in Sec. IV. Cutoff-energy dependence of the
calculated energies is demonstrated in these sections. Summary follows in Sec. V.

\section{$G$-matrix including reduced NN force from 3NF}
It is difficult to treat the 3NF $V_{123}$ directly in infinite matter.
In this paper, we introduce an approximation of reducing the 3NF to an effective
NN force by folding single-nucleon degrees of freedom, as has been
often employed in the literature \cite{KAT74,FP81,LNR71,HKW10}. That is,
the density dependent NN interaction $V_{12(3)}$ is defined,
in momentum space, by the following summation over the third nucleon in
the Fermi sea of nuclear matter:
\begin{widetext}
\begin{equation}
 \langle \bk_1' \sigma_1'\tau_1',\bk_2' \sigma_2'\tau_2'|V_{12(3)}| \bk_1 \sigma_1\tau_1,
 \bk_2 \sigma_2\tau_2\rangle_A
 \equiv \sum_{\bk_3,\sigma_3 \tau_3} \langle \bk_1' \sigma_1'\tau_1',
 \bk_2' \sigma_2'\tau_2', \bk_3 \sigma_3\tau_3|V_{123}| \bk_1 \sigma_1\tau_1,
 \bk_2 \sigma_2\tau_2, \bk_3 \sigma_3\tau_3\rangle_A.
\end{equation}
\end{widetext}
The suffix $A$ denotes an antisymmetrized matrix element, namely
$|ab\rangle_A \equiv |ab-ba\rangle$ and $|abc\rangle_A \equiv |abc-acb+bca-bac+cab-cba\rangle$.
The remaining two nucleons are supposed to be in a center-of-mass frame;
$\bk_1'+\bk_2'=\bk_1+\bk_2=0$.
We do not include the three-body form factor in this folding procedure, but introduce it
later in the reduced NN interaction. In this case, matrix elements and their partial wave
expansion can be carried out analytically for the Ch-EFT 3NF, as presented in Appendix A.

The necessity of taking into account of correlations being neglected, contributions of the
two- and three-nucleon forces, $V_{12}$ and $V_{123}$, to the energy are given by
\begin{eqnarray}
& & \frac{1}{2} \sum_{\bk_1 \bk_2} \langle \bk_1 \bk_2 | V_{12} |\bk_1 \bk_2\rangle_A \nonumber \\
& &+ \frac{1}{3!}\sum_{\bk_1 \bk_2 \bk_3} \langle \bk_1 \bk_2 \bk_3| V_{123} |\bk_1 \bk_2 \bk_3\rangle_A
 \nonumber \\
&=& \frac{1}{2} \sum_{\bk_1 \bk_2} \langle \bk_1 \bk_2 | V_{12}+\frac{1}{3} V_{12(3)} |\bk_1 \bk_2\rangle_A
\end{eqnarray}
This implies that the $G$-matrix may be defined by
\begin{equation}
 G_{12}=V_{12}+\frac{1}{3}V_{12(3)}+(V_{12}+\frac{1}{3}V_{12(3)})\frac{Q}{\omega - H}G_{12},
\end{equation}
where $Q$ stands for the Pauli exclusion operator and the denominator $\omega -H$ of the
propagator is prescribed below. The similar evaluation of the single-particle energy needs
a different combination factor:
\begin{eqnarray}
 && \langle \bk |t|\bk\rangle + \sum_{\bk'} \langle \bk \bk' | V_{12} |\bk \bk' \rangle_A \nonumber \\
 & & + \frac{1}{2} \sum_{\bk' \bk''} \langle \bk \bk' \bk'' | V_{123} |\bk \bk'\bk''\rangle_A \nonumber \\
 &=& \langle \bk |t|\bk\rangle+\sum_{\bk'} \langle \bk \bk' | V_{12}+\frac{1}{2} V_{12(3)} |\bk \bk' \rangle_A ,
\end{eqnarray}
where $t$ is a kinetic-energy operator. It is reasonable to define the single-particle
energy which is used in the denominator of the $G$-matrix equation, Eq. (3),
employing the continuous prescription for intermediate states as
\begin{eqnarray}
 e_{\bk} &=& \langle \bk |t|\bk\rangle + U_G(\bk) \\
 U_G(\bk) &\equiv & \sum_{\bk'} \langle \bk \bk' | G_{12} \nonumber \\
 & & +\frac{1}{6} V_{12(3)}\left(1+\frac{Q}{\omega - H}\right)G_{12} |\bk \bk' \rangle_A ,
\end{eqnarray}
supposing that effects of the NN correlation is approximated by that of the
$G$-matrix equation. To be specific, the denominator $\omega -H$ in the $G$-matrix
equation for $G|\bk_1 \bk_2\rangle$ is given by
$e_{\bk_1}+e_{\bk_2}-(t_1+U_G(\bk_1')+t_2+U_G(\bk_2'))$, where $\bk_1'$ and $\bk_2'$ are
momenta of intermediate nucleons.

Solving the $G$-matrix equation together with the denominator explained above,
the total energy is evaluated by:
\begin{eqnarray}
 E &=& \sum_{\bk} \langle \bk |t|\bk\rangle + \frac{1}{2}\sum_{\bk} U_E(\bk) \\
 U_E(\bk) &=& \sum_{\bk'} \langle \bk \bk' | G_{12} |\bk \bk'\rangle_A
\end{eqnarray}
The difference between $U_G(\bk)$ for the energy calculation and $U_E(\bk)$
appeared in the single-particle energy is a prototype of rearrangement energy. 
Naturally, the above treatment of the 3NF is heuristic. It is desirable
to develop a more rigorous and systematic perturbative treatment. One possible
framework may be a coupled cluster method, which was discussed in Ref. \cite{KO12}. 

In actual calculations in nuclear matter, a partial wave expansion \cite{HT70} is introduced with
an angle-average approximation for the Pauli exclusion operator $Q$.
The good quality of this approximation has been examined in the literature \cite{SOKN00}.
The partial wave expansion of the reduced NN interaction, Eq. (1), is carried out
in a standard way, which may be found in the paper by Fujiwara {\it et al.} \cite{YF00}.
Partial waves up to the total angular momentum $J=7$ and the orbital angular
momentum $\ell=7$ are included in numerical calculations.

For completeness, explicit expressions of the reduced NN interactions of $V_C$, $V_D$,
and $V_E$ parts and their partial wave contributions are given in Appendices A and B.
Similar calculations were presented by Holt, Kaiser and Weise \cite{HKW10}. We, however,
do not use an approximation for the off-diagonal matrix elements assumed there. 
It is possible to obtain analytical expressions for the partial wave expansion
by introducing several functions in a form of the integration of Legendre polynomials
of the second kind, as given in Eqs. (B1)-(B6).
All terms in $V_C$ and $V_D$ yield central and tensor interactions. Spin-orbit components appear
only in the $c_1$ and $c_3$ terms of $V_C$. The $V_E$ interaction gives only an $\ell =0$ central
component; that is, in the $^1S_0$ and $^3S_1$ channels.

Low-energy constants of the Ch-EFT interaction used in numerical calculations in the
following sections are those of the J\"{u}lich group \cite{EGM05}:
$c_1=-0.81$ GeV$^{-1}$, $c_3=-3.4$ GeV$^{-1}$,
and $c_4=3.4$ GeV$^{-1}$. Other constants are taken from the Ref. \cite{HEB11}:
$c_D=-4.381$ and $c_E=-1.126$. As noted in Appendix A, the reduced effective
interaction $V_{12(3)}$ is multiplied by a form factor $exp\{-(q'/\Lambda)^6-(q/\Lambda)^6\}$.
We assume the same cutoff $\Lambda$ as in the NN sector. 

\section{Numerical calculations in symmetric nuclear matter}
First, we present results of LOBT calculations in symmetric nuclear matter,
using only the initial NN part of the Ch-EFT potential.
It is expected that the obtained saturation curve is not much different from
those of other modern NN potentials. The Ch-EFT potential is regularized by
a rather soft form factor as the interaction based on low-energy effective theory.
The nuclear-matter energy may depend considerably on the cutoff energy
$\Lambda$ of the regulator. We show in the beginning the results with
$\Lambda=550$ MeV, and later discuss the $\Lambda$-dependence.
The obtained saturation curve in symmetric nuclear matter is shown by
a dashed curve in Fig. 1, compared with results of other NN potentials:
AV18 \cite{AV18}, NSC \cite{NSC}, and CD-Bonn \cite{CDB} potentials. It is seen
that the very similar saturation curve to those of AV18 and NSC is obtained. AV18 and
NSC have comparatively stronger tensor component than CD-Bonn, which is
reflected in the larger deuteron D-state probability. Although the Ch-EFT interaction
shows a smaller deuteron D-state probability, the LOBT energy is similar to those
of AV18 and NSC.

\begin{figure}
\includegraphics[width=0.45\textwidth]{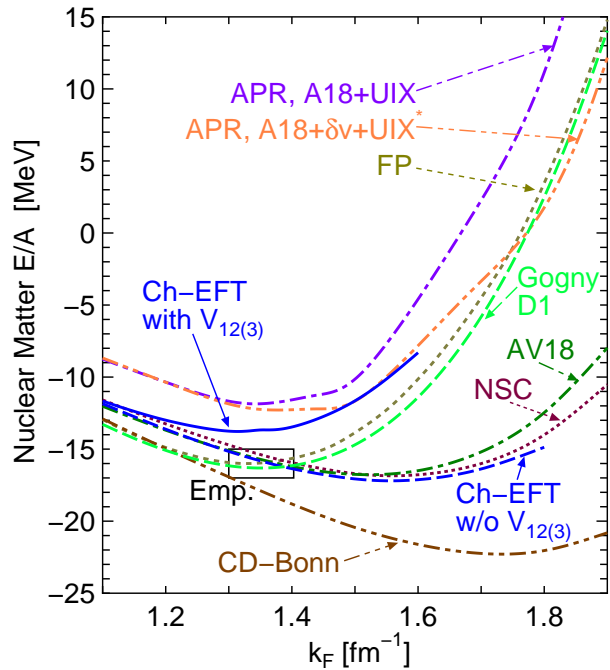}
\caption{
LOBT saturation curves in symmetric nuclear matter,
using the Ch-EFT interaction with the cutoff energy of $\Lambda=$550 MeV.
The solid and dashed curves are the results with and without the 3NF effects,
respectively. Energies from other modern NN potentials,
AV18 \cite{AV18}, NSC \cite{NSC} and CD-Bonn \cite{CDB}, are also shown.
As a basis for comparison, the energy expected from the Gogny D1 force \cite{GP77}
is included. In addition, results of variational calculations by the Illinois group
with the phenomenological 3NF, FP \cite{FP81} and APR \cite{APR98}, are included.
}
\end{figure}

When the effect of the 3NF is included by the procedure explained in Sect. 2,
we obtain the solid curve shown in Fig. 1. As a basis for comparison, the saturation
curve expected from the Gogny force \cite{GP77}, which is an standard effective
interaction used for a density-dependent Hartree-Fock description of nuclei,
is also plotted. The calculation at higher densities than $k_F=1.6$ fm$^{-1}$ is
unreliable and not shown, because calculated s.p. energies wobble badly at large
momentum beyond the normal density where the Ch-EFT as the low-energy
theory is not to be applied especially when the reduce NN force is included. The
saturation property is much improved by including $V_{12(3)}$, although the
energy at the saturation point is shallow by a few MeV. The deviation is
probably within the uncertainty of the lowest-order calculation on the one
hand, and the uncertainties of low-energy constants as well as the ambiguity
of cut-off parameters on the other. Therefore, the long-standing problem
of microscopic understanding of the nuclear saturation seems to be resolved
by the inclusion of the 3NF. This recognition may not be new, because the
role of the 3NF has been demonstrated repeatedly in the literature \cite{KAT74,FP81,APR98}.
However, previous calculations inevitably include phenomenological adjustment.
The advantage of the present calculation with using the Ch-EFT 3NF
interaction is that the potential is systematically constructed and is
consistent with the NN sector. The $c_3$ term of the Ch-EFT
3NF is found to give dominant repulsive contribution to the energy.
This coupling constant is determined in the NN sector and
therefore no room for an additional adjustment.

\begin{figure}[t]
\includegraphics[width=0.40\textwidth]{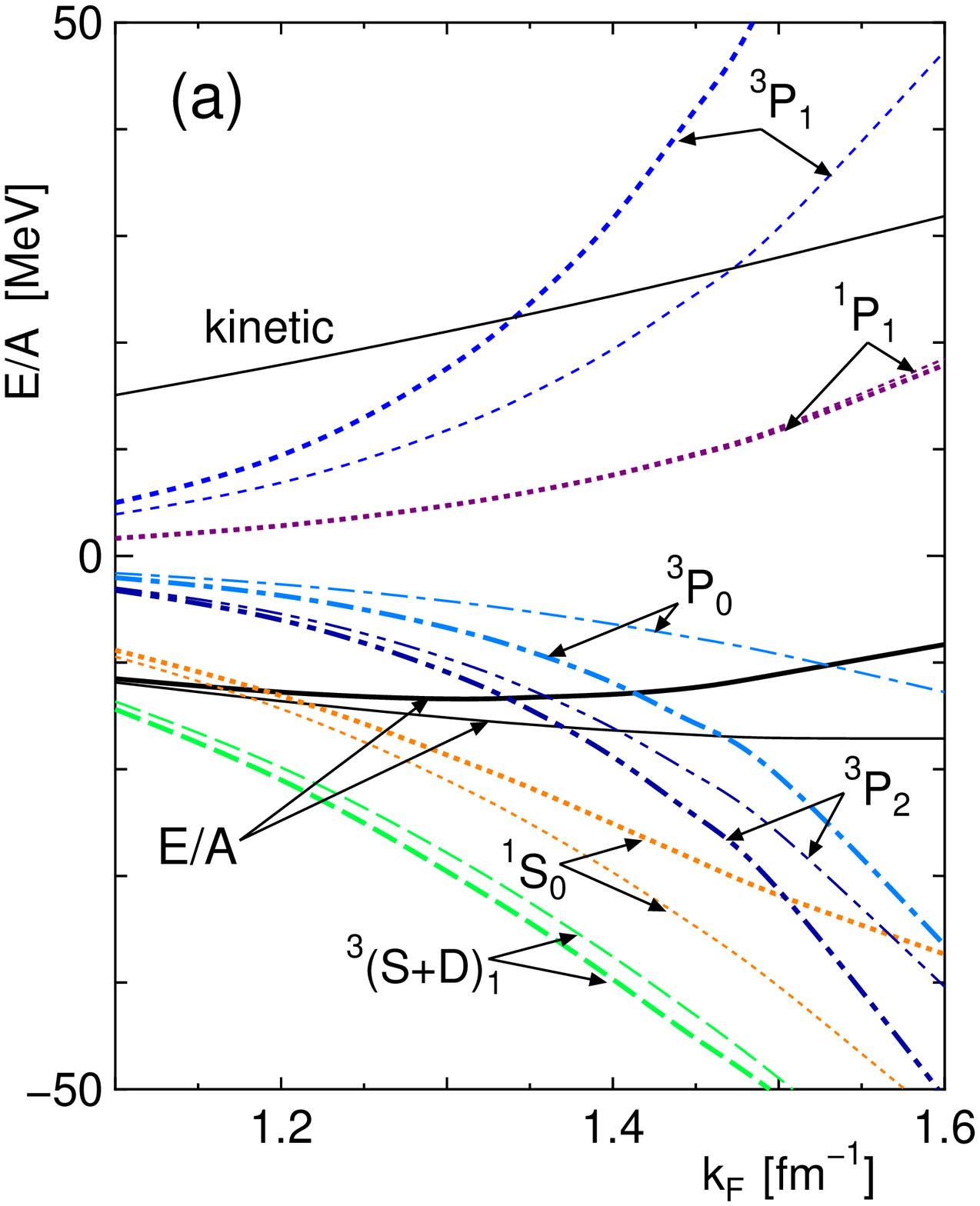}
\includegraphics[width=0.40\textwidth]{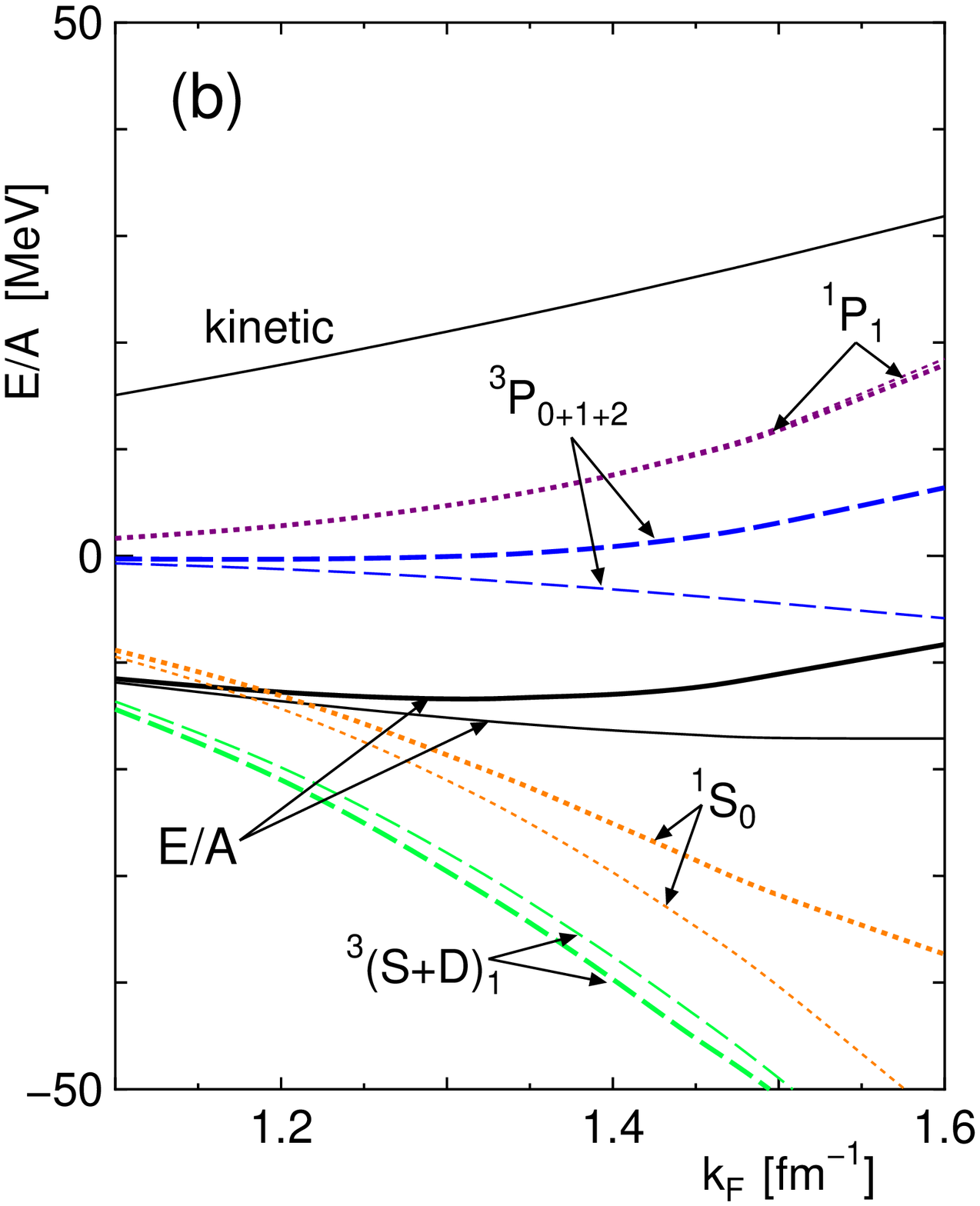}
\caption{
$k_F$-dependence of partial wave contributions
to the nuclear matter LOBT energy per nucleon for the Ch-EFT interaction
with $\Lambda=550$ MeV. Thick and thin curves are with and without the 3NF
effects, respectively; (a) full decomposition, (b) different $J$ being summed.
}
\end{figure}

To see the contributions of the 3NF in more details, we show, in Fig. 2,
partial wave decomposition of the calculated potential energy.
The attractive contribution in the $^3S_1$ channel is seen to increase
by including the 3NF. This is due to the enhancement
of the tensor correlation by the supplemented tensor force.
On the other hand, the $^1S_0$ contribution becomes less attractive.
The $p$-wave contributions depend much on the total-angular
momentum $J$. This is owing to the rather strong spin-orbit component.
It has been known that the net effect of the triplet $p$-wave contribution is small,
which is a rather remarkable character of the NN interaction. This property
persists after including the 3NF, but the net $^3$O contribution
becomes repulsive when the 3NF is incorporated. The repulsion gradually grows 
as the density goes up, is important for improving the description of the nuclear
saturation property. On the other hand, the singlet $p$ channel is affected little
by the 3NF. These characteristics of the 3NF contributions may be utilized for
improving the effective interactions for density-dependent Hartree-Fock
calculations and/or density functionals for medium-heavy nuclei.

It has been recognized in nuclear structure calculations that the
two-body spin-orbit force is not sufficient to explain
a strong single-particle spin-orbit field which is essential to describe empirical
nuclear shell structures characterized by nuclear magic numbers.
As was shown in the separate paper \cite{MK12}, the additional spin-orbit
strength from the 3NF is favorable to provide the empirical strength of the
one-body spin-orbit field. The strength of the nuclear one-body
spin-orbit potential from nucleon-nucleon interactions is represented
by the Scheerbaum factor $B_S(\bar{q})$, the definition of which is found
in Ref. \cite{MK12}.  $B_S(\bar{q})$ corresponds to the spin-orbit strength $W$
of the $\delta$-type two-body spin-orbit interaction
$iW(\bfsigma_1+\bfsigma_2)\cdot [\nabla_r\times (\br)\nabla_r]$ customarily
used in nuclear Hartree-Fock calculations. The empirical value of $W$ is
around 120 MeV$\cdot$fm$^5$. Because those results in Ref. \cite{MK12} were
simply obtained by $G_{12}$ and not by $G_{12}+\frac{1}{6} V_{12(3)}
\left(1+\frac{Q}{\omega -H}\right) G_{12}$ explained in Sec. II, we show
revised numbers in Table I. The additional term makes the value of $B_S(\bar{q})$
slightly larger. 

\begin{table}[b]
\begin{tabular}{crrrr}\hline \hline
                              & \multicolumn{2}{c}{nuclear matter} & \multicolumn{2}{c}{neutron matter} \\ 
 $k_F=1.35$ fm$^{-1}$ &  N$^3$LO & N$^3$LO+3NF &  N$^3$LO & N$^3$LO+3NF \\ \hline
 $B_S(T=0)$ &  2.5 & 7.3 & -- & -- \\
 $B_S(T=1)$ & 84.6 & 120.2 & 84.7 & 93.3 \\ \hline \hline
                              & \multicolumn{2}{c}{nuclear matter} & \multicolumn{2}{c}{neutron matter} \\ 
$k_F=1.07$ fm$^{-1}$ & N$^3$LO & N$^3$LO+3NF &  N$^3$LO & N$^3$LO+3NF \\ \hline
 $B_S(T=0)$ & 1.6 & 4.4 & -- & -- \\
 $B_S(T=1)$ & 86.5 & 109.8 & 87.0 & 92.3 \\ \hline \hline
\end{tabular}
\caption{Scheerbaum factor $B_S(\bar{q})$ in units of MeV$\cdot$fm$^5$
with $\bar{q}=0.7$ fm$^{-1}$ for J\"{u}lich N$^3$LO \cite{EHM09}
with and without 3NF. The $G$-matrix in Ref. \cite{MK12} is replaced by
$G_{12}+\frac{1}{6}V_{12(3)}\left(1+\frac{Q}{\omega-H}\right)G_{12}$ in
this calculation.
}
\end{table}

\begin{figure}
\includegraphics[width=0.45\textwidth]{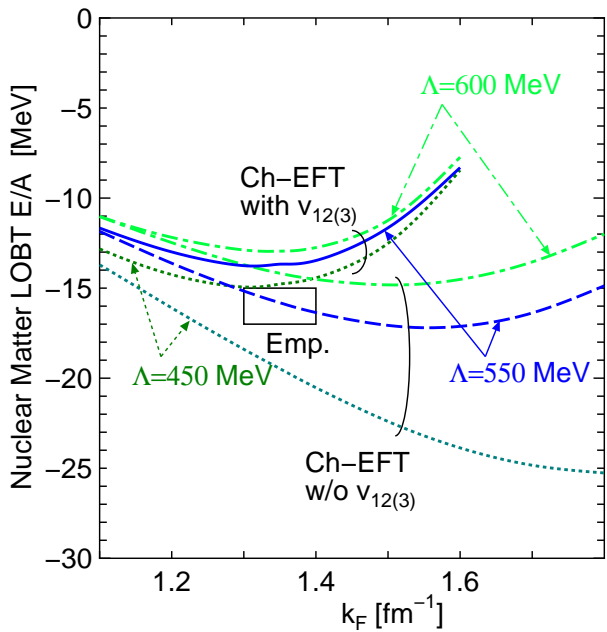}
\caption{
Cutoff $\Lambda$-dependence of the LOBT energy per nucleon
in symmetric nuclear matter for the Ch-EFT interaction with and without the 3NF effects.
}
\end{figure}

Now, we address the cutoff-energy dependence of calculated LOBT energies.
We show, in Fig. 3, saturation curves with using $\Lambda=450$ MeV and
$\Lambda=600$ MeV, in addition to the case of $\Lambda=550$ MeV presented in Fig. 1.
When only the NN interactions are employed, the calculated energies
depend considerably on $\Lambda$. The smaller cutoff energy provides larger
binding-energies. The result with $\Lambda=450$ MeV is rather close to that of the
CD-Bonn potential given in Fig. 1. It is impressive to observe that calculated
energies with different $\Lambda$ become very close each other when the
3NF is added. That is, the cutoff-energy dependence is significantly reduced
if the NN and 3NF which are constructed consistently in the Ch-EFT are
simultaneously used in the LOBT calculation.

\begin{figure}
\includegraphics[width=0.45\textwidth]{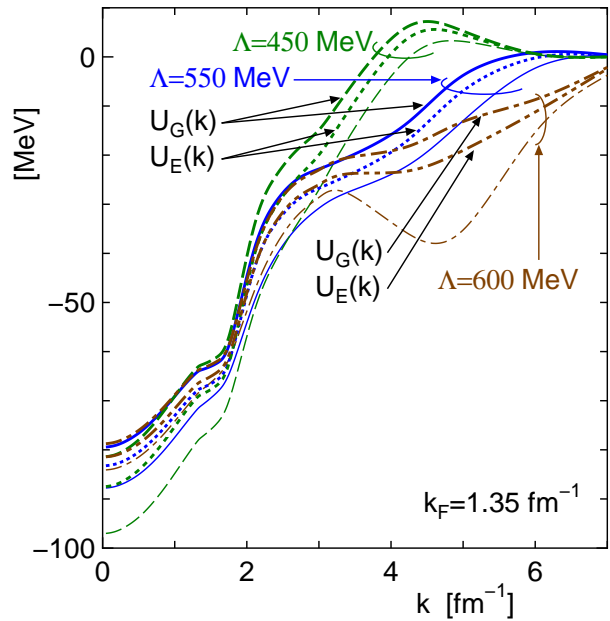}
\caption{
Momentum dependence of single-particle energies
$U_G(k)$, Eq. (6), and $U_E(k)$, Eq. (8),
in symmetric nuclear matter for the three cases of the cutoff energy $\Lambda$
of the Ch-EFT interaction. Thin curves show the results without the 3NF effects. 
}
\end{figure}

Finally in this section, we remark on the quantitative difference between $U_G(\bk)$
and $U_E(\bk)$ defined in Eqs. (6) and (8), respectively. Figure 4 compares $U_G(\bk)$
and $U_E(\bk)$ with the NN force and 3NF for three cases of the cutoff energy $\Lambda$.
The s.p. potential $U_E(\bk)$ without the 3NF effects is also shown. The difference
of $U_G(\bk)$ and $U_E(\bk)$, which
is $\sum_{\bk'} \frac{1}{6} \langle \bk\bk'V_{12(3)} \left(1+\frac{Q}{\omega -H}\right)
 G_{12} |\bk\bk'\rangle_A$, is on the order of 5 MeV for $|\bk| \leq 2$ fm$^{-1}$.
That is, the s.p. energy is raised by around 5 MeV
by the additional term. Through the starting energy dependence of
the $G$-matrix, the total energy per nucleon is lowered by about 0.5 MeV.
The large $\Lambda$-dependence of the s.p. potential beyond $|\bk|= 3$
fm$^{-1}$ has no physical significance. As the results in Fig. 3 suggest, $U_E(\bk)$
for $|\bk| \ltsim 2$ fm$^{-1}$ does not depend much on the cutoff energy,
when the 3NF is included.
 
\section{Numerical calculations in pure neutron matter}
The energy per nucleon of neutron matter is fundamental to determine properties of
neutron star matter. The $k_F$ dependence of calculated LOBT energies
in pure neutron matter with and without including the 3NF is shown in Fig 5.
Energies obtained with other modern NN potentials and results of the variational
calculation by the Illinois group \cite{FP81,APR98} are also presented for comparison.
The latter used the AV18 potential \cite{AV18} and included the
3NF of the Fujita-Miyazawa \cite{FM57} type supplemented by phenomenological
terms. Because the strong tensor effect in the $^3S_0$-$^3D_0$ channel
is absent, many-body correlations are relatively simple in neutron matter.
Since the calculated saturation curve in symmetric nuclear matter already
well corresponds to the empirical one, the present LOBT energy in neutron matter
is expected to be trustful. In contrast to the symmetric nuclear matter, calculated energies
with different NN potentials are very similar, as is seen in the $k_F$-dependence of
neutron matter energies with Ch-EFT, AV18, NSC and CD-Bonn potentials in Fig. 5.

The Ch-EFT 3NF itself is more predictive for the application to neutron matter,
because the contact $c_E$ term vanishes in pure neutron matter as
the Pauli principle forbids three neutrons to assemble at the same place,
and the $c_D$ term which gives null in the plane wave case gives a negligibly
small contribution. In addition, the $c_4$ term does not contribute.
Thus the contribution from the NNLO 3NF is determined by the $c_1$
and $c_3$ terms. These coupling constants are determined in the
NN sector.

\begin{figure}
\includegraphics[width=0.45\textwidth]{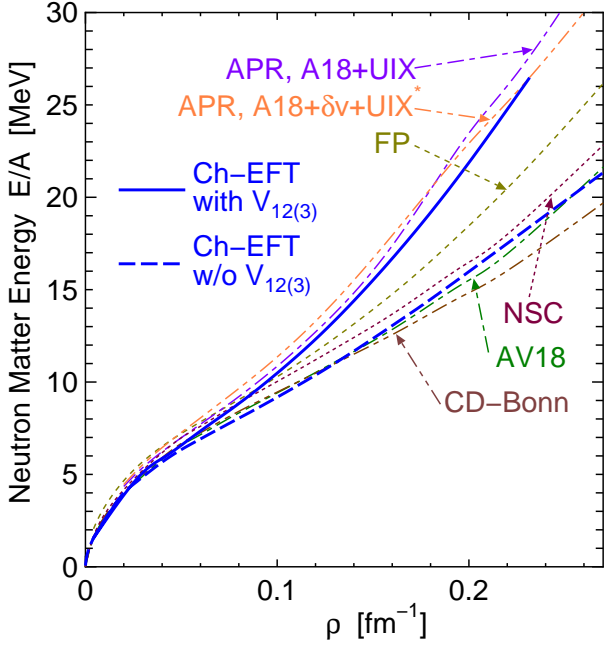}
\caption{
Calculated LOBT energy per nucleon in pure neutron matter,
using the Ch-EFT interaction with the cutoff energy of $\Lambda=$550 MeV with
and without including effects of the 3NF. Energies from other modern NN potentials,
AV18 \cite{AV18}, NSC \cite{NSC}, and CD-Bonn \cite{CDB}, and results of
variational calculations by the Illinois group, FP \cite{FP81} and APR \cite{APR98},
are also shown.
}
\end{figure}

The results of the variational calculation in Ref. \cite{APR98} shown in Fig. 5 have
been utilized as the canonical equation of state (EoS) for discussing
neutron star matter properties. It is interesting to see that the LOBT result obtained
with including the 3NF, in which no phenomenological adjustment is introduced,
is close to the EoS of Ref. \cite{APR98}.

As the Ch-EFT cannot be applied to high momentum region, it is not
possible to discuss directly the EoS relevant to the core of high-density neutron stars.
However, it is possible to provide the reference EoS at lower densities which
should be smoothly matched to the EoS obtained by theories designed for
the high density region. Such an attempt was recently reported in Ref. \cite{S13}.

As was noted in Ref. \cite{MK12}, the magnitude of the spin-orbit
component in $V_{12(3)}$ obtained in pure neutron matter is one third
of that in symmetric nuclear matter. Although correlations somewhat modifies
this number, as is given in Table I, the calculated additional
contribution to the Scheerbaum factor from the 3NF in neutron matter
is about $\frac{1}{3}$ of that in nuclear matter. Observing that the
contribution of the genuine NN interaction to the s.p. spin-orbit strength is insensitive to
the neutron-proton asymmetry $\alpha=\frac{N-Z}{N+Z}$, the 3NF can be the
source of the asymmetry dependence of the strength of the s.p. spin-orbit potential.
In a Woods-Saxon potential model, rather strong $\alpha$-dependence,
such as $(1-0.54\alpha)$ was inferred, as in the textbook by Bohr-Mottelson \cite{BM}.
Recent fitting \cite{YRS} gives gentler $\alpha$-dependence, typically $1-0.25\alpha$.
If we naively use the calculated numbers given in Table I and assume that $B_S(T=1)$
in neutron matter depends little on the density, the $\alpha$-dependence of the s.p.
spin-orbit strength is estimated as $(1-0.22\alpha)$, which is consistent with the value
mentioned above.

\begin{figure}
\includegraphics[width=0.45\textwidth]{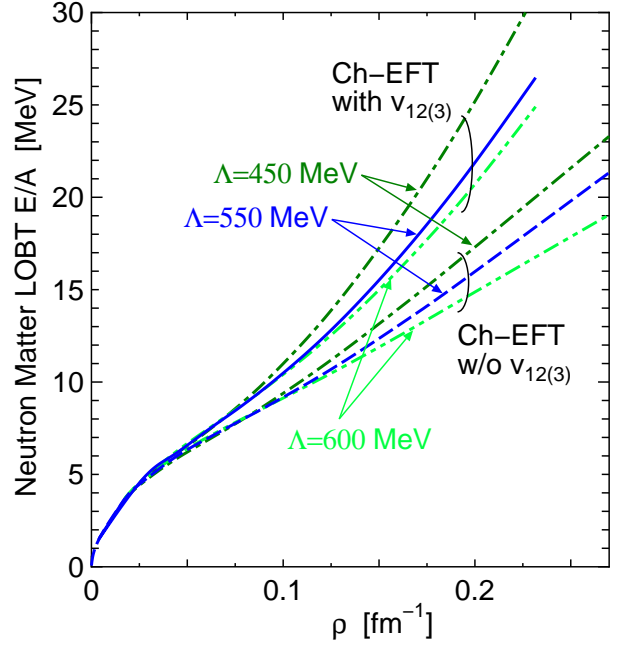}
\caption{
Cutoff $\Lambda$-dependence of the LOBT energy per nucleon
in pure neutron matter for the Ch-EFT interaction with and without the 3NF effects.
}
\end{figure}

Finally, Fig. 6 shows the variation in the neutron matter energy for a different choice
of the cutoff energy $\Lambda$. The $\Lambda$-dependence of the calculated energies
is already moderate in the case of the NN interaction only.

\section{Summary}
We have calculated LOBT energies both in symmetric nuclear matter and
pure neutron matter, using the Ch-EFT N$^3$LO NN interaction and NNLO
3NF of the J\"{u}lich group \cite{EGM05}. In the Ch-EFT, the 3NF is
introduced in a systematic way along with the NN potential.
Three of  5 coupling constants in the NNLO 3NF are fixed in a NN
sector. The remaining two parameters are under control in the literature
to reproduce properties of few-nucleon systems.
The 3NF is treated by reducing it to density-dependent NN interactions
by folding single-nucleon degrees of freedom in infinite matter.
We have given, in the Appendices, explicit expressions of the reduced
NN interactions and their partial-wave expanded forms.

Calculated results show that the empirical saturation property is well
reproduced in nuclear matter. In a conventional understanding, effects
of the Pauli blocking for the strong tensor coupling have been emphasized
as the basic mechanism of causing the nuclear saturation property. Though
this effect is fundamentally important, the sizable repulsive contribution
of the 3NF is also crucial in the region around and above the normal nuclear
matter density. This indicates that the Pauli blocking not only for the standard
tensor correlation but also for other non-nucleonic excitations inherent
in the two-nucleon process, such as isobar $\Delta$ and anti-nucleon
excitations, provides large repulsive effects.

It is noteworthy that the large cutoff-energy dependence of
calculated energies obtained only with the Ch-EFT NN force reduces substantially
when including the 3NF effects. This dependence arises predominantly
in the triplet even channel; that is, the channel in which the tensor correlation
is significant. Therefore, the cutoff-energy dependence is rather weak in neutron matter.

Contributions of the 3NF in the $^1E$ and $^3O$ channels are repulsive.
Owing to the repulsion, the density-dependence of neutron matter energy
per nucleon becomes very close to those favorable for describing neutron
star properties, although the prediction of the Ch-EFT cannot be applied
at high densities. The strength of the spin-orbit component in the $^3O$
channel increases by about 30\%, which resolves the problem of the insufficiency
of modern NN potentials to account for the empirical spin-orbit strength,
as previously reported in Ref. \cite{MK12}. The potential energy in the $^3S_1$ state
turns out to become more attractive due to the enhancement of the
tensor component. The knowledge of these specific properties of the
3NF contributions may be helpful for improving
effective forces and/or energy-functionals for finite nuclei.

We conclude that although more rigorous treatment of the 3NF together
with more than three-body correlations are required in future, the present
calculations demonstrate that the 3NF constructed consistently with
the NN part in the sense of effective field theory can reproduce
basic nuclear properties, namely the saturation and strong spin-orbit field,
without phenomenological adjustments.

\acknowledgments
This work is supported by JSPS KAKENHI Grant Numbers 22540288 and 25400266.
The author thanks H. Kamada for valuable comments concerning the Ch-EFT
interactions. He is also grateful to M. Yahiro for his interest in this work.

\newpage
\appendix
\begin{widetext}
\section{Effective NN forces from the 3NF in chiral effective field theory}
In the leading order, NNLO, three-nucleon force $V_{123}$ consists of
terms specified by five low-energy coupling constants $c_1$, $c_3$, $c_4$, $c_D$, and $c_E$:
$V_{123}=V_C+V_D+V_E$. Each term is given as follows.
\begin{eqnarray}
V_C &=& \frac{1}{2} \left( \frac{g_A}{2f_\pi}\right)^2 \sum_{i\ne j\ne k}
 \frac{(\bfsigma_i \cdot \bq_i)(\bfsigma_j \cdot \bq_j)}{(q_i^2+m_\pi^2)(q_j^2+m_\pi^2)}
 \tau_i^\alpha \tau_j^\beta \left\{ \delta^{\alpha\beta}
\left[ - \frac{4c_1 m_\pi^2}{f_\pi^2}+\frac{2c_3}{f_\pi^2}\bq_i\cdot \bq_j \right] \right.\nonumber \\
 & &\left.  +\sum_\gamma \frac{c_4}{f_\pi^2} \epsilon^{\alpha\beta\gamma}\tau_k^\gamma
 \bfsigma_k \cdot (\bq_i\times \bq_j) \right\}, \\
 V_D &=& -\frac{g_A}{8f_\pi^2} \frac{c_D}{f_\pi^2 \Lambda_\chi} \sum_{i\ne j\ne k}
 \frac{(\bfsigma_j\cdot \bq_j)(\bfsigma_i \cdot \bq_j)}{q_j^2+m_\pi^2} (\bftau_i\cdot \bftau_j), \\
 V_E &=& \frac{c_E}{2f_\pi^4\Lambda_\chi} \sum_{j\ne k} (\bftau_j\cdot \bftau_k )
 = \frac{c_E}{f_\pi^4\Lambda_\chi} (\bftau_1\cdot \bftau_2 +\bftau_2\cdot \bftau_3+
\bftau_3\cdot \bftau_1 ).
\end{eqnarray}
The three coupling constants $c_1$, $c_3$, and $c_3$ are determined in the NN
sector and the remaining $c_D$ and $c_E$ are adjusted in more than three nucleon systems.
As is explained in Eq. (1), the three-nucleon force $V_{123}$ is reduced to an effective
NN force $V_{12(3)}$ by summing over the third nucleon in the Fermi sea:
\begin{equation}
 \langle \bk_1' \sigma_1'\tau_1',-\bk_1' \sigma_2'\tau_2'|V_{12(3)}| \bk_1 \sigma_1\tau_1,
 -\bk_1 \sigma_2\tau_2\rangle_a
 \equiv \sum_{\bk_3,\sigma_3 \tau_3} \langle \bk_1' \sigma_1'\tau_1',
 -\bk_1' \sigma_2'\tau_2', \bk_3 \sigma_3\tau_3|V_{123}| \bk_1 \sigma_1\tau_1,
 -\bk_1 \sigma_2\tau_2, \bk_3 \sigma_3\tau_3\rangle_a .
\end{equation}
Form factors are not taken into account in this folding procedure. The obtained $V_{12(3)}$
is multiplied by a form factor in the form of $exp\{-(q'/\Lambda)^6-(q/\Lambda)^6\}$.
 
In this appendix, we present details of the matrix elements of $\langle \bk_1'\sigma_1'\tau_1',
-\bk_1'\sigma_2'\tau_2'|V_{12(3)}|\bk_1\sigma_1\tau_1,-\bk_1\sigma_2\tau_2\rangle$ from
Ch-EFT 3NF forces $V_C$, $V_D$, and $V_E$. In the following, we use the standard notation
for the tensor operator $S_{12}(\bk',\bk)$ and their matrix elements between partial waves:
\begin{eqnarray}
 & & S_{12}(\bk',\bk)=3 ([\bfsigma_1\times\bfsigma_2]^2\cdot [\bk'\times\bk]^2),\\
 & & \left(S_{12}\right)^{\ell'}_{\ell 1 J}=\frac{6\sqrt{J(J+1)}}{2J+1} \;\mbox{for }\;\ell'=\ell+2\;\mbox{or}\;
 \ell'=\ell-2, \\
 & & \left(S_{12}\right)^{J}_{J 1 J}=2,\;\; \left(S_{12}\right)^{J+1}_{J+1 1 J}=-\frac{2(J+2)}{2J+1},\;\;
 \mbox{and}\;\;\left(S_{12}\right)^{J-1}_{J-1 1 J}=-\frac{2(J-1)}{2J+1}.
\end{eqnarray}
Evaluating Eq. (A4), the $c_1$ term of $V_C$ provides
\begin{eqnarray}
 & & \frac{c_1 g_A^2 m_\pi^2 \rho_0}{f_\pi^4}
  \frac{(\bfsigma_1\cdot (\bk_1'-\bk_1))(\bfsigma_2\cdot(\bk_1'-\bk_1))}
 {((\bk_1'-\bk_1)^2+m_\pi^2)((\bk_1'-\bk_1)^2+m_\pi^2)}(\bftau_1\cdot\bftau_2) \nonumber \\
  & & +\frac{c_1 g_A^2 m_\pi^2}{f_\pi^4}\sum_{\bk_3}
 \left(\frac{\frac{2}{3} (\bfsigma_1\cdot\bfsigma_2)(\bk_1'-\bk_1)\cdot (\bk_3+\bk_1)
 +2([\bfsigma_1 \times \bfsigma_2]^2\cdot [ (\bk_1'-\bk_1)\times (\bk_3+\bk_1)]^2)}
 {((\bk_1'-\bk_1)^2+m_\pi^2)((\bk_3+\bk_1)^2+m_\pi^2)} \right. \nonumber \\
 & &  \hspace*{20mm}\left. +\frac{\frac{2}{3} (\bfsigma_1\cdot\bfsigma_2)(\bk_1'-\bk_3)\cdot (\bk_1-\bk_1')
 +2([\bfsigma_1\times \bfsigma_2]^2 \cdot [ (\bk_1'-\bk_3)\times (\bk_1-\bk_1')]^2)}
 {((\bk_1'-\bk_3)^2+m_\pi^2)((\bk_1'-\bk_1)^2+m_\pi^2)} \right) (\bftau_1\cdot\bftau_2) \nonumber \\
 & & +\frac{c_1 g_A^2 m_\pi^2}{f_\pi^4}\sum_{\bk_3}\frac{6(\bk_1'-\bk_3)\cdot(\bk_3-\bk_1)
 +3i(\bfsigma_1+\bfsigma_2)\cdot((\bk_1'-\bk_3)\times(\bk_3-\bk_1))}
 {((\bk_1'-\bk_3)^2+m_\pi^2)((\bk_3-\bk_1)^2+m_\pi^2)},
\end{eqnarray}
the $c_3$ term of $V_C$
\begin{eqnarray}
 & &  \frac{c_3 g_A^2\rho_0}{2f_\pi^4} \frac{(\bfsigma_1\cdot (\bk_1'-\bk_1))(\bfsigma_2\cdot (\bk_1'-\bk_1))}
 {((\bk_1'-\bk_1)^2+m_\pi^2)^2} |\bk_1'-\bk_1|^2(\bftau_1\cdot\bftau_2)  \nonumber \\
  & & -\frac{c_3 g_A^2}{2f_\pi^4}\frac{\frac{2}{3}(\bftau_1\cdot\bftau_2)
 (\bfsigma_1\cdot\bfsigma_2)}{(\bk_1'-\bk_1)^2+m_\pi^2}
  \left\{((\bk_1'-\bk_1)\cdot \bk_1)^2 (F_0(k_1)-2F_1(k_1)) \right. \nonumber \\
 & & +((\bk_1'-\bk_1)\cdot \bk_1')^2 (F_0(k_1')-2 F_1(k_1'))
  +\frac{1}{3} |\bk_1'-\bk_1|^2 (k_1^2 F_{2}(k_1)+k_1'^2 F_{2}(k_1'))\nonumber\\
 & & \left. +\frac{1}{k_1^2} ([(\bk_1'-\bk_1)\times (\bk_1'-\bk_1)]^2\cdot
 [\bk_1\times \bk_1]^2)k_1^2F_{3}(k_1) 
  +\frac{1}{k_1^2} ([(\bk_1'-\bk_1)\times (\bk_1'-\bk_1)]^2\cdot [\bk_1'\times \bk_1']^2)k_1'^2F_{3}(k_1')
 \right\} \nonumber \\
 & & +\frac{c_3 g_A^2}{2f_\pi^4}\frac{2(\bftau_1\cdot\bftau_2)}{(\bk_1'-\bk_1)^2+m_\pi^2}
  \left\{ \frac{1}{3} S_{12}(\bk_1'-\bk_1, \bk_1)
  ((\bk_1'-\bk_1)\cdot\bk_1)(F_0(k_1)-2F_1(k_1)) \right. \nonumber\\
 & & +\frac{1}{3} S_{12}(\bk_1'-\bk_1, \bk_1') ((\bk_1'-\bk_1)\cdot\bk_1')(F_0(k_1')-2F_1(k_1'))
  +\frac{1}{9} S_{12}(\bk_1'-\bk_1, \bk_1'-\bk_1) (k_1^2 F_{2}(k_1)+k_1'^2 F_{2}(k_1')) \nonumber\\
 & &-\frac{1}{9}[ S_{12}(\bk_1,\bk_1)(-2k_1^2+3(\bk_1\cdot \bk_1'))+S_{12}(\bk_1',\bk_1') k_1^2
 +S_{12}(\bk_1',\bk_1) (k_1^2-3(\bk_1\cdot\bk_1'))]F_{3}(k_1) \nonumber\\
 & &\left. -\frac{1}{9}[ S_{12}(\bk_1',\bk_1')(-2k_1'^2+3(\bk_1\cdot \bk_1'))+S_{12}(\bk_1,\bk_1) k_1'^2
 +S_{12}(\bk_1',\bk_1) (k_1'^2-3(\bk_1\cdot\bk_1'))]F_{3}(k_1') \right\}\nonumber\\
  & &  + \frac{c_3 g_A^2}{2f_\pi^4}\sum_{\bk_3} \frac{6(\bk_1'-\bk_3)\cdot(\bk_3-\bk_1)
 +3i(\bfsigma_1+\bfsigma_2)\cdot((\bk_1'-\bk_1)\times\bk_3-\bk_1'\times \bk_1)}
 {((\bk_1'-\bk_3)^2+m_\pi^2)((\bk_3-\bk_1)^2+m_\pi^2)} (\bk_1'-\bk_3)\cdot (\bk_3-\bk_1),
\end{eqnarray}
and the $c_4$ term of $V_C$
\begin{eqnarray}
 & & 2\frac{c_4 g_A^2}{4f_\pi^4} \sum_{\bk_3} \left\{
  \frac{(\bfsigma_1\cdot (\bk_1'-\bk_1))(\bfsigma_2\cdot ((-\bk_1'-\bk_3)\times
((\bk_1'-\bk_1)\times (-\bk_1'-\bk_3)))}
{((\bk_1'-\bk_1)^2+m_\pi^2)((-\bk_1'-\bk_3)^2+m_\pi^2)} \right. \nonumber\\
  & &  + \frac{(\bfsigma_1\cdot (\bk_1'-\bk_1))
(\bfsigma_2\cdot (((\bk_3+\bk_1)\times(\bk_1'-\bk_1))\times(\bk_3+\bk_1))}
{((\bk_1'-\bk_1)^2+m_\pi^2)((\bk_3+\bk_1)^2+m_\pi^2)}\nonumber\\
  & &  +\frac{(\bfsigma_1\cdot ((\bk_1'-\bk_3)\times(\bk_1-\bk_3)))
(\bfsigma_2\cdot((\bk_1'-\bk_3)\times(\bk_1-\bk_3)))}
{((\bk_1'-\bk_3)^2+m_\pi^2)((\bk_1-\bk_3)^2+m_\pi^2)}\nonumber\\
 & &- \frac{(\bfsigma_1\cdot ((\bk_1'-\bk_3)\times((\bk_1'-\bk_3)\times(-\bk_1'+\bk_1))))
(\bfsigma_2\cdot(-\bk_1'+\bk_1))}
{((\bk_1'-\bk_3)^2+m_\pi^2)((\bk_1'-\bk_1)^2+m_\pi^2)}\nonumber\\
  & & -\frac{(\bfsigma_1\cdot ((\bk_1'-\bk_3)\times(\bk_3-\bk_1))
 (\bfsigma_2\cdot (\bk_3-\bk_1)\times(\bk_1'-\bk_3))}
 {((\bk_1'-\bk_3)^2+m_\pi^2)((\bk_3-\bk_1)^2+m_\pi^2)}\nonumber\\
 & & \left. -\frac{(\bfsigma_1\cdot (((-\bk_1'+\bk_1)
 \times(\bk_3-\bk_1))\times(\bk_3-\bk_1))(\bfsigma_2\cdot (-\bk_1'+\bk_1))}
{((-\bk_1'+\bk_1)^2+m_\pi^2)((\bk_3-\bk_1)^2+m_\pi^2)} \right\}
 (\bftau_1\cdot\bftau_2).
\end{eqnarray}

The $V_D$ term is found to yield
\begin{eqnarray}
& & -\frac{g_A}{8f_\pi^2} \frac{c_D \rho_0}{f_\pi^2 \Lambda_\chi}
 \frac{\frac{1}{3}(\bfsigma_1\cdot \bfsigma_2)(\bk_1'-\bk_1)^2+([\bfsigma_1\times \bfsigma_2]^2
 \cdot [(\bk_{1}'-\bk_1)\times (\bk_1'-\bk_1)]^2)}{(\bk_1'-\bk_1)^2+m_\pi^2}
 (\bftau_1\cdot \bftau_2) \nonumber\\
 & & +2\frac{g_A}{8f_\pi^2} \frac{c_D}{f_\pi^2 \Lambda_\chi} \left\{
 \frac{1}{3}(\bfsigma_1\cdot\bfsigma_2)
\left(\frac{1}{2}\rho_0-m_\pi^2 F_0(k_1)-m_\pi^2 F_0(k_1')\right)\right.\nonumber\\
 & &  + ([\bfsigma_1\times \bfsigma_2]^2 \cdot [(\bk_{1}'-\bk_1)\times (\bk_1'-\bk_1)]^2)
  \left( F_0(k_1)-2F_1(k_1)+F_3(k_1)+F_0(k_1')-2F_1(k_1')+F_3(k_1')\right)\}
  (\bftau_1\cdot \bftau_2) \nonumber\\
 & & +6\frac{g_A}{8f_\pi^2} \frac{c_D}{f_\pi^2 \Lambda_\chi}
 \left\{\frac{1}{2}\rho_0-m_\pi^2F_0(k_1)-m_\pi^2F_0(k_1')\right\}.
\end{eqnarray}

Finally, the $V_E$ term gives a spin- and isospin-scalar interaction:
\begin{equation}
-6\frac{C_E\frac{1}{4}\rho_0}{f_\pi^4\Lambda_\chi}.
\end{equation}
In the above expressions, (A9) and (A11), functions $F_0$, $F_1$, $F_2$, and $F_4$ are defined
as follows.
\begin{eqnarray}
 F_0(k) & \equiv & \frac{1}{(2\pi)^3} \iiint_{|\bk'| \leq  k_F} d\bk' \frac{1}{(\bk-\bk')^2+m_\pi^2} \nonumber \\
  &=& \frac{1}{(2\pi)^2} \left\{ k_F+\frac{k_F^2+m_\pi^2-k^2}{4k}
 \log\frac{(k+k')^2+m_\pi^2}{(k-k')^2+m_\pi^2}-m_\pi \left(\arctan \frac{k+k_F}{m_\pi}
 -\arctan \frac{k-k_F}{m_\pi} \right)  \right\}, \\
 F_1(k) &\equiv & \frac{1}{k}\frac{1}{(2\pi)^3}\iiint_{|\bk'| \leq  k_F} d\bk'
 \frac{k'\cos\theta }{(\bk-\bk')^2+m_\pi^2}\nonumber \\
 &=& \frac{1}{k}\frac{1}{(2\pi)^2} \left[ \frac{k_F}{4k}(3k^2-k_F^2-m_\pi^2)
 -km_\pi \left(\arctan \frac{k+k_F}{m_\pi}-\arctan\frac{k-k_F}{m_\pi}\right) \right.\nonumber \\
 & & \left. +\frac{1}{16k^2} \{m_\pi^4+2m_\pi^2(3k^2+k_F^2)+(k_F^2-k^2)(k_F^2+3k^2)\}
 \log \frac{(k+k_F)^2+m_\pi^2}{(k-k_F)^2+m_\pi^2}\right], \\
  F_{2}(k)&\equiv& \frac{2\pi}{(2\pi)^3} \frac{1}{k^3} \int_0^{k_F}dk' k'^3
 Q_0\left( \frac{k^2+k'^2+m_\pi^2}{2kk'}\right)\nonumber \\
 &=& \frac{1}{(2\pi)^2}\frac{1}{k^2} \left\{ \frac{1}{6} k_F(3k^2+k_F^2-9m_\pi^2)
 +\frac{(k_F^4-k^4-m_\pi^4+6k^2m_\pi^2)}{8k} \log \frac{(k+k_F)^2+m_\pi^2}{(k-k_F)^2+m_\pi^2} \right.\nonumber\\
 & & \left. +m_\pi (m_\pi^2-k^2) \left(\arctan \frac{k+k_F}{m_\pi}-\arctan\frac{k-k_F}{m_\pi}\right) \right\},\\
 F_{3}(k)&\equiv& \frac{2\pi}{(2\pi)^3} \frac{1}{k^3} \int_0^{k_F}dk' k'^3
 Q_2\left( \frac{k^2+k'^2+m_\pi^2}{2kk'}\right)\nonumber\\
  &=& \frac{1}{k^2}\frac{1}{(2\pi)^2} \left\{ \frac{1}{32k^3} [(k_F^2+k^2+m_\pi^2)^3
  +2k^2m_\pi^2(m_\pi^2+6k^2)-2k^2(k_F^4+3k^4)] \right. \nonumber\\
  & & \times \log \frac{(k+k_F)^2+m_\pi^2}{(k-k_F)^2+m_\pi^2}
  -k^2m_\pi \left(\arctan \frac{k+k_F}{m_\pi}-\arctan\frac{k-k_F}{m_\pi}\right) \nonumber\\
  & & \left. -\frac{k_F}{8k^2}\left (m_\pi^4+4k^2m_\pi^2+k_F^4-5k^4+\frac{4}{3}k^2k_F^2
 +2m_\pi^2k_F^2 \right) \right\}.
\end{eqnarray}
\section{Partial wave expansion}
The expressions of the Born kernel $\langle \bk_1'\sigma_1'\tau_1',
-\bk_1'\sigma_2'\tau_2'|V_{12(3)}|\bk_1\sigma_1\tau_1,-\bk_1\sigma_2\tau_2\rangle$
in the previous section need to be expanded into partial waves
for standard nuclear-matter $G$-matrix calculations. The procedure may be found in Ref. \cite{YF00}.
We use the abbreviated notations for integrals involving second kind Legendre
functions $Q_{\ell}$'s.
\begin{eqnarray}
 & & Q_{W0}^\ell(k_1',k_1)\equiv \frac{1}{(2\pi)^2}\frac{1}{2} \int_0^{k_F} dk_3 Q_{\ell}(x')Q_{\ell}(x),\\
 & & Q_{W2}^\ell(k_1',k_1)\equiv \frac{1}{(2\pi)^2}\frac{1}{2k_1'k_1} \int_0^{k_F} dk_3 k_3^2Q_{\ell}(x')Q_{\ell}(x),\\
 & & Q_{W4}^\ell(k_1',k_1)\equiv \frac{1}{(2\pi)^2}\frac{1}{2(k_1'k_1)^2} \int_0^{k_F} dk_3 k_3^4Q_{\ell}(x')Q_{\ell}(x),\\
 & & Q_{W1}^\ell(k_1',k_1)\equiv \frac{1}{(2\pi)^2}\frac{1}{2k_1} \int_0^{k_F} dk_3 k_3x'Q_{\ell}(x')Q_{\ell}(x),\\
 & & Q_{W1}^\ell(k_1,k_1')\equiv \frac{1}{(2\pi)^2}\frac{1}{2k_1'} \int_0^{k_F} dk_3 k_3xQ_{\ell}(x')Q_{\ell}(x),
\end{eqnarray}
where $x\equiv \frac{k_1^2+k_3^2+m_\pi^2}{2k_1k_3}$ and $x'\equiv \frac{k_1'^2+k_3^2+m_\pi^2}{2k_1'k_3}$.

The central component of the $c_1$ interaction of $V_C$, Eq. (A8), with an orbital angular momentum $\ell$ is
\begin{eqnarray}
 & &  \frac{c_1 g_A^2 m_\pi^2 \rho_0}{f_\pi^4}  \frac{1}{3}
 (\bfsigma_1\cdot \bfsigma_2)(\bftau_1\cdot\bftau_2)\left(\frac{1}{2k_1'k_1}Q_\ell(z)
 +\frac{m_\pi^2}{(2k_1'k_1)^2}{Q_\ell}'(z)\right) \nonumber\\
  & &  -\frac{2}{3}\frac{c_1 g_A^2 m_\pi^2}{f_\pi^4}
 (\bfsigma_1\cdot \bfsigma_2)(\bftau_1\cdot\bftau_2)\left\{\frac{1}{2k_1'k_1}Q_\ell(z)
  (\bk_1'^2(F_0(k_1')-F_1(k_1'))+\bk_1^2(F_0(k_1)-F_1(k_1))) \right.\nonumber\\
 & & \left. - \frac{1}{2}Q_\ell^{(1)}(z) (F_0(k_1')+F_0(k_1)-F_1(k_1')-F_1(k_1))\right\}
  -6 \frac{c_1 g_A^2 m_\pi^2}{f_\pi^4}\left( \delta_{\ell 0} \frac{1}{2}(F_0(k_1')+F_0(k_1))\right.\nonumber\\
 & & \left. + \frac{\ell+1}{2\ell+1}Q_{W0}^{\ell+1} (k_1',k_1)+\frac{\ell}{2\ell+1} Q_{W0}^{\ell-1} (k_1',k_1)
  +Q_{W2}^\ell (k_1',k_1)-Q_{W1}^\ell (k_1',k_1) -Q_{W1}^\ell (k_1,k_1') \right).
\end{eqnarray}
The tensor components of the $c_1$ interaction of $V_C$, Eq. (A8), are
\begin{eqnarray}
 & & \frac{c_1 g_A^2 m_\pi^2 \rho_0}{f_\pi^4}  \frac{1}{3}(\bftau_1\cdot\bftau_2) 
 (S_{12})_{\ell 1J}^{\ell'} \frac{-1}{2k_1'k_1}\left(k_1'^2 \frac{1}{2k_1'k_1}{Q_\ell}' (z) +k_1^2 \frac{1}{2k_1'k_1}{Q_{\ell'}}' (z)
 -{Q_J}' (z) \right) \nonumber\\
  & & -\frac{2}{3}\frac{c_1 g_A^2 m_\pi^2}{f_\pi^4}
  (\bftau_1\cdot\bftau_2)(S_{12})_{\ell 1J}^{\ell'} \left( \frac{k_1^2}{2k_1'k_1}Q_{\ell'}(z)(F_0(k_1')-F_1(k_1'))\right.\nonumber\\
 & & \left.  +\frac{k_1'^2}{2k_1'k_1}Q_{\ell}(z)(F_0(k_1)-F_1(k_1)) -\frac{1}{2} Q_J(z)(F_0(k_1')+F_0(k_1)-F_1(k_1')-F_1(k_1))\right)
\end{eqnarray}
for $\ell'=\ell\pm 1$ ($J=\ell\pm 1$) and
\begin{eqnarray}
 & & \frac{c_1 g_A^2 m_\pi^2 \rho_0}{f_\pi^4}  \frac{1}{3}(\bftau_1\cdot\bftau_2) 
 (S_{12})_{\ell 1J}^{\ell'} \frac{-1}{2k_1'k_1}
 \left\{k_1'^2 \frac{1}{2k_1'k_1}{Q_\ell}' (z) +k_1^2 \frac{1}{2k_1'k_1}{Q_{\ell'}}' (z)\right. \nonumber\\
 & & \hspace*{5cm} \left. -\frac{1}{2}\left(\frac{2\ell+3}{2\ell+1}{Q_{\ell-1}}'(z)
 +\frac{2\ell-1}{2\ell+1} {Q_{\ell+1}}'(z)\right)\right\} \nonumber\\
  & & -\frac{2}{3}\frac{c_1 g_A^2 m_\pi^2}{f_\pi^4}
  (\bftau_1\cdot\bftau_2)(S_{12})_{\ell 1J}^{\ell'} \left\{ \frac{k_1^2}{2k_1'k_1}Q_{\ell'}(z)(F_0(k_1')-F_1(k_1'))
 +\frac{k_1'^2}{2k_1'k_1}Q_{\ell}(z)(F_0(k_1)-F_1(k_1))\right. \nonumber\\
 & & \left.  -\frac{1}{2}\left(\frac{2\ell+3}{2\ell+1}{Q_{\ell-1}}(z)+\frac{2\ell-1}{2\ell+1} {Q_{\ell+1}}(z)
 \right)(F_0(k_1')+F_0(k_1)-F_1(k_1')-F_1(k_1))\right\}.
\end{eqnarray}
for $\ell'=\ell=J\pm 1$. The spin-orbit component of the $c_1$ term of $V_C$, Eq. (A8), becomes
\begin{equation}
\delta_{S1} \frac{c_1g_A^2m_{\pi}^2}{f_{\pi}^4} 3\frac{\ell(\ell+1)+2-J(J+1)}{2\ell+1}
\{ -Q_{W0}^{\ell-1}(k_1',k_1)+Q_{W0}^{\ell+1}(k_1',k_1)+W_{\ell s,0}^{\ell}(k_1',k_1) \},
\end{equation}
where the function $W_{\ell s,0}^{\ell}(k_1',k_1)$ is defined as
\begin{equation}
 W_{\ell s,0}^{\ell}(k_1',k_1)=\frac{4\pi}{(2\pi)^3} \int_0^\infty dk_3 \frac{k_3}{4k_1'k_1}
 \{k_1'Q_{\ell}(x) (Q_{\ell-1}(x')-Q_{\ell+1}(x'))+k_1Q_{\ell}(x) (Q_{\ell-1}(x)-Q_{\ell+1}(x))\}.
\end{equation}

The central component of the $c_3$ interaction of $V_C$, Eq. (A9), is
\begin{eqnarray}
&&  \frac{c_3 g_A^2\rho_0}{2f_\pi^4} \frac{1}{3}(\bfsigma_1\cdot \bfsigma_2)(\bftau_1\cdot\bftau_2)
 \left\{ \delta_{\ell 0} -\frac{m_\pi^2}{k_1'k_1}Q_\ell (z)+\left(\frac{m_\pi^2}{2k_1'k_1}\right)^2
 \frac{\ell+1}{z^2-1}(zQ_\ell (z)-Q_{\ell+1}(z)) \right\} \nonumber\\
  & & -\frac{c_3 g_A^2}{2f_\pi^4} \frac{2}{3}(\bftau_1\cdot\bftau_2) (\bfsigma_1\cdot\bfsigma_2)
  \left\{ \frac{1}{3} \left[ \delta_{\ell 0} -m_\pi^2 \frac{1}{2k_1'k_1} Q_\ell(z)\right] ( k_1^2 F_{2}(k_1)+ k_1'^2 F_{2}(k_1'))
\right.  \nonumber\\
 & & +\left[-\frac{1}{2}k_1'k_1 \frac{1}{3}\delta_{\ell 1} -\frac{1}{4} (k_1'^2-3k_1^2+m_\pi^2)\delta_{\ell 0}
  + \frac{1}{4}(k_1'^2-k_1^2+m_\pi^2)^2 \frac{1}{2k_1'k_1} Q_\ell(z)\right] (F_0(k_1)-2F_1(k_1))  \nonumber \\
 & & +\left[ -\frac{1}{2}k_1'k_1 \frac{1}{3}\delta_{\ell 1} -\frac{1}{4} (k_1^2-3k_1'^2+m_\pi^2)\delta_{\ell 0}
 + \frac{1}{4}(k_1^2-k_1'^2+m_\pi^2)^2 \frac{1}{2k_1'k_1} Q_\ell(z)\right] (F_0(k_1')-2F_1(k_1')) \nonumber\\
 & & +\left[ -\frac{k_1'^2}{6k_1'k_1}\delta_{\ell 1}
-\frac{1}{2k_1'k_1}\left[ \frac{k_1'^2}{2k_1'k_1}(k_1'^2+k_1^2+m_\pi^2)-\frac{4}{3}k_1'k_1 \right] \delta_{\ell 0}
 \right. \nonumber \\
 & & \left. +\left[ \left(\frac{k_1'}{2k_1'k_1}\right)^2(k_1'^2+k_1^2+m_\pi^2)^2-k_1'^2
 -\frac{2}{3}m_\pi^2\right]\frac{1}{2k_1'k_1} Q_\ell(z)\right] k_1^2F_{3}(k_1) \nonumber\\
 & & +\left[ -\frac{k_1^2}{6k_1'k_1}\delta_{\ell 1}
-\frac{1}{2k_1'k_1}\left[ \frac{k_1^2}{2k_1'k_1}(k_1'^2+k_1^2+m_\pi^2)-\frac{4}{3}k_1'k_1 \right] \delta_{\ell 0}
 \right. \nonumber\\
 & & \left. +\left[ \left(\frac{k_1}{2k_1'k_1}\right)^2(k_1'^2+k_1^2+m_\pi^2)^2-k_1^2
 -\frac{2}{3}m_\pi^2\right]\frac{1}{2k_1'k_1} Q_\ell(z)\right]k_1'^2 F_{3}(k_1') \nonumber\\
  & &  - \frac{c_3 g_A^2}{2f_\pi^4}6
  \left\{ \delta_{\ell 0}\left[\frac{1}{8}\rho_0-\left(\frac{3}{4}m_\pi^2 +\frac{1}{2}k_1'^2+\frac{1}{4}k_1^2\right)F_0(k_1')
 -\left(\frac{3}{4}m_\pi^2 +\frac{1}{2}k_1^2+\frac{1}{4}k_1'^2\right)F_0(k_1)\right. \right. \nonumber\\
 & & \left. +\frac{1}{4}(k_1'^2 F_2(k_1')+k_1^2 F_2(k_1)) \right]+ \delta_{\ell 1} \frac{k_1'k_1}{3} \left[ F_0(k_1')+F_0(k_1)-
 \frac{1}{2}(F_1(k_1')+F_1(k_1))\right]  \nonumber\\
 & &  +\frac{1}{4k_1'k_1}(k_1'^2+k_1^2+2m_\pi^2)^2 Q_{W0}^\ell(k_1',k_1)-(k_1'^2+k_1^2+2m_\pi^2)
 \left( \frac{\ell}{\hat{\ell}} Q_{W0}^{\ell-1}(k_1',k_1)+ \frac{(\ell+1)}{\hat{\ell}} Q_{W0}^{\ell+1}(k_1,k_1')\right) \nonumber\\
 & & \left. + \frac{k_1'k_1}{\hat{\ell}} \left[ \frac{(\ell+1)(\ell+2)}{2\ell+3}Q_{W0}^{\ell+2}(k_1',k_1)
 +\left(\frac{\ell^2}{2\ell-1}+\frac{(\ell+1)^2}{2\ell+3} \right) Q_{W0}^\ell (k_1',k_1)
  +\frac{\ell(\ell-1)}{2\ell-1}Q_{W0}^{\ell-2}(k_1',k_1)\right] \right\}.
\end{eqnarray}
The tensor components of the $c_3$ interaction of $V_C$, Eq. (A8), are
 \begin{eqnarray}
 & &  \frac{c_3 g_A^2}{2f_\pi^4}  (S_{12})_{\ell 1J}^{\ell'}\frac{\rho_0}{3} (\bftau_1\cdot\bftau_2) \left[
  \frac{k_1'^2}{2k_1'k_1}\left(Q_\ell (z)+\frac{m_\pi^2}{2k_1'k_1}Q_{\ell}'(z)\right) \right. \nonumber \\
  & & \hspace{7cm} \left.  +\frac{k_1^2}{2k_1'k_1}\left(Q_{\ell'} (z) +\frac{m_\pi^2}{2k_1'k_1}Q_{\ell'}'(z)\right)
  -\left( Q_J(z)+\frac{m_\pi^2}{2k_1'k_1}Q_{J}'(z)\right) \right] \nonumber\\
 & & -\frac{c_3 g_A^2}{2f_\pi^4} \frac{2}{3} (\bftau_1\cdot\bftau_2)\left[ (F_0(k_1)-2F_1(k_1))
 \left\{-k_1^2 \left(\frac{1}{2}Q_{\ell'}^{(1)}(z) -\frac{k_1^2}{2k_1'k_1}Q_{\ell'}(z) \right)
 \right. + k_1'k_1\left(\frac{1}{2}Q_{J}^{(1)}(z) -\frac{k_1^2}{2k_1'k_1}Q_{J}(z) \right) \right\} \nonumber \\
 & & +(F_0(k_1')-2F_1(k_1')) \left\{k_1'^2 \left(-\frac{1}{2}Q_{\ell}^{(1)}(z)+\frac{k_1'^2}{2k_1'k_1}Q_{\ell}(z)\right)
  -k_1'k_1\left(-\frac{1}{2}Q_{J}^{(1)}(z) +\frac{k_1'^2}{2k_1'k_1}Q_{J}(z) \right) \right\} \nonumber\\
 & & +\frac{1}{3}(k_1'^2 F_2(k_1')+k_1^2 F_2(k_1)) \left\{ \frac{k_1'^2}{2k_1'k_1}Q_{\ell}(z)+\frac{k_1^2}{2k_1'k_1}Q_{\ell'}(z)
 -Q_J(z)\right\} \nonumber \\
 & & +\frac{1}{3}(2k_1^2F_3(k_1)-k_1'^2F_3(k_1')) \frac{k_1^2}{2k_1'k_1}Q_{\ell'}(z)
 -\frac{1}{2}k_1^2 F_3(k_1)Q_{\ell'}^{(1)}(z)  \nonumber\\
 & & +\frac{1}{3}(2k_1'^2F_3(k_1')-k_1^2F_3(k_1)) \frac{k_1'^2}{2k_1'k_1}Q_{\ell}(z)
 -\frac{1}{2}k_1'^2 F_3(k_1')Q_{\ell}^{(1)}(z) \nonumber \\
 & & \left. - \frac{1}{3}(k_1^2F_3(k_1)+k_1'^2F_3(k_1')) \frac{1}{2}Q_{J}(z)
 +\frac{1}{2}k_1'k_1(F_3(k_1')+F_3(k_1)) Q_{J}^{(1)}(z) \right]
 \end{eqnarray}
for $\ell'=\ell\pm 1$ ($J=\ell\pm 1$) and
 \begin{eqnarray}
  & &  \frac{c_3 g_A^2}{2f_\pi^4}(S_{12})_{\ell 1J}^{\ell'} \frac{\rho_0}{3} (\bftau_1\cdot\bftau_2) \left[
  \frac{k_1'^2+k_1^2}{2k_1'k_1}\left(Q_\ell (z)+\frac{m_\pi^2}{2k_1'k_1}Q_{\ell}'(z)\right) \right.  \nonumber\\
 & & \left. - \frac{1}{2} \left\{ \frac{2\ell+3}{2\ell+1}\left( Q_{\ell-1}(z)+\frac{m_\pi^2}{2k_1'k_1}Q_{\ell-1}'(z)\right)
 +\frac{2\ell-1}{2\ell+1}\left( Q_{\ell+1}(z)+\frac{m_\pi^2}{2k_1'k_1}Q_{\ell+1}'(z)\right) \right\}\right]  \nonumber\\
  & & -\frac{c_3 g_A^2}{2f_\pi^4} \frac{2}{3} (\bftau_1\cdot\bftau_2)\left[ (F_0(k_1)-2F_1(k_1))
 \left\{-k_1^2 \left(\frac{1}{2}Q_{\ell}^{(1)}(z) -\frac{k_1^2}{2k_1'k_1}Q_{\ell}(z) \right) \right. \right.  \nonumber\\
 & & + \frac{1}{2} k_1'k_1\left\{ \frac{2\ell+3}{2\ell+1}\left(\frac{1}{2}Q_{\ell-1}^{(1)}(z)
 -\frac{k_1^2}{2k_1'k_1}Q_{\ell-1}(z) \right) +\frac{2\ell-1}{2\ell+1}\left(\frac{1}{2}Q_{\ell+1}^{(1)}(z)
 -\frac{k_1^2}{2k_1'k_1}Q_{\ell+1}(z) \right)\right\}  \nonumber\\
 & &  +(F_0(k_1')-2F_1(k_1')) \left\{k_1'^2 \left(-\frac{1}{2}Q_{\ell}^{(1)}(z)
 +\frac{k_1'^2}{2k_1'k_1}Q_{\ell}(z) \right) \right. \nonumber\\
 & & + \frac{1}{2} k_1'k_1\left\{ \frac{2\ell+3}{2\ell+1}\left(\frac{1}{2}Q_{\ell-1}^{(1)}(z)
 -\frac{k_1'^2}{2k_1'k_1}Q_{\ell-1}(z) \right) +\frac{2\ell-1}{2\ell+1}\left(\frac{1}{2}Q_{\ell+1}^{(1)}(z)
 -\frac{k_1'^2}{2k_1'k_1}Q_{\ell+1}(z) \right)\right\}  \nonumber\\
 & & +\frac{1}{3}(2k_1^2F_3(k_1)-k_1'^2F_3(k_1')) \frac{k_1^2}{2k_1'k_1}Q_{\ell}(z)
 -\frac{1}{2}k_1^2 F_3(k_1)Q_{\ell}^{(1)}(z)  \nonumber\\
 & & +\frac{1}{3}(2k_1'^2F_3(k_1')-k_1^2F_3(k_1)) \frac{k_1'^2}{2k_1'k_1}Q_{\ell}(z)
 -\frac{1}{2}k_1'^2 F_3(k_1')Q_{\ell}^{(1)}(z) \nonumber \\
 & & - \frac{1}{3}(k_1^2F_3(k_1)+k_1'^2F_3(k_1')) \frac{1}{4}\left\{\frac{2\ell +3}{2\ell+1}Q_{\ell-1}(z)
 +\frac{2\ell -1}{2\ell+1}Q_{\ell+1}(z)\right\} \nonumber\\
 & & \left. +k_1'k_1(F_3(k_1')+F_3(k_1')) \frac{1}{4}\left\{\frac{2\ell +3}{2\ell+1}Q_{\ell-1}^{(1)}(z)
 +\frac{2\ell -1}{2\ell+1}Q_{\ell+1}^{(1)}(z)\right\}\right]
\end{eqnarray}
 for $\ell'=\ell=J\pm 1$. The spin-orbit component of the $c_3$ term of $V_C$, Eq. (A9), becomes
\begin{eqnarray}
 & & \delta_{S1} \frac{c_3g_A^2}{2f_{\pi}^4} 3\frac{\ell(\ell+1)+2-J(J+1)}{2\ell+1} \left[
 \left( m_{\pi}^2+\frac{1}{2}(k_1'^2+k_1^2)\right)
\{ Q_{W0}^{\ell-1}(k_1',k_1)-Q_{W0}^{\ell+1}(k_1',k_1)-W_{\ell s,0}^{\ell}(k_1',k_1) \} \right. \nonumber \\
 & & \left. -\delta_{\ell 1} \frac{k_1'k_1}{2} (F_0(k_1')+F_0(k_1)-F_1(k_1')-F_1(K_1))
 + 6\frac{k_1'k_1}{2} \left\{ \frac{\ell-1}{2\ell-1} W_{\ell s,0}^{\ell-1}(k_1',k_1)
 +\frac{\ell+2}{2\ell+3} W_{\ell s,0}^{\ell+1}(k_1',k_1)\right\} \right].
\end{eqnarray}

The central component of the $c_4$ interaction of $V_C$, Eq. (A10), is
\begin{eqnarray}
 & & 2\frac{c_4 g_A^2}{4f_\pi^4} \frac{2}{3}(\bfsigma_1\cdot\bfsigma_2) (\bftau_1\cdot\bftau_2)
 \left[\left(\frac{1}{2}\rho_0-m_\pi^2 (F_0(k_1')+F_0(k_1))-\frac{1}{3}(k_1'^2 F_2(k_1')+k_1^2 F_2(k_1))\right)
 \left(\delta_{\ell 0}-\frac{m_\pi^2}{2k_1'k_1}Q_\ell (z)\right) \right.  \nonumber\\
 & & +(F_0(k_1')-2F_1(k_1'))\left( \frac{1}{4}(k_1^2-3k_1'^2+m_\pi^2)\delta_{\ell 0}
 +\frac{1}{6}k_1'k_1 \delta_{\ell 1}-\frac{1}{4}(k_1^2-k_1'^2+m_\pi^2)^2 \frac{1}{2k_1'k_1}Q_\ell (z) \right) \nonumber\\
 & & +(F_0(k_1)-2F_1(k_1))\left( \frac{1}{4}(k_1'^2-3k_1^2+m_\pi^2)\delta_{\ell 0}
 +\frac{1}{6}k_1'k_1 \delta_{\ell 1}-\frac{1}{4}(k_1'^2-k_1^2+m_\pi^2)^2 \frac{1}{2k_1'k_1}Q_\ell (z) \right) \nonumber\\
 & & +\left[\left\{ \frac{1}{4} (k_1'^2+k_1^2+m_\pi^2)-\frac{2}{3}k_1'^2\right\}\delta_{\ell 0}
 +\frac{1}{6}k_1'k_1 \delta_{\ell 1}-\left\{ \frac{1}{4}  (k_1'^2+k_1^2+m_\pi^2)^2
 -k_1'^2k_1^2-\frac{2}{3}m_\pi^2 k_1'^2 \right\} \frac{1}{2k_1'k_1}Q_\ell (z)\right] F_3(k_1')  \nonumber\\
 & & +\left[\left\{ \frac{1}{4} (k_1'^2+k_1^2+m_\pi^2)-\frac{2}{3}k_1^2\right\}\delta_{\ell 0}
 +\frac{1}{6}k_1'k_1 \delta_{\ell 1} -\left\{ \frac{1}{4}  (k_1'^2+k_1^2+m_\pi^2)^2
 -k_1'^2k_1^2-\frac{2}{3}m_\pi^2 k_1^2 \right\} \frac{1}{2k_1'k_1}Q_\ell (z)\right] F_3(k_1)  \nonumber\\
 & & + \delta_{\ell 0} \left\{
 +\frac{1}{8}\rho_0+\frac{1}{4}(2k_1'^2+k_1^2-m_\pi^2)F_0(k_1')+\frac{1}{4}(k_1'^2+2k_1^2-m_\pi^2)F_0(k_1)
-\frac{1}{4}(k_1'^2 F_2(k_1')+k_1^2 F_2(k_1))\right\} \nonumber\\
 & & +\delta_{\ell 1} k_1'k_1\left\{\frac{1}{6}(F_1(k_1')+F_1(k_2))-\frac{1}{3}(F_0(k_1')+F_0(k_1))\right\}
 - \frac{1}{4}(k_1'^2+k_1^2)(k_1'^2+k_1^2+4m_\pi^2) \frac{1}{k_1'k_1}Q_{W0}^\ell (k_1',k_1) \nonumber\\
  & & +(k_1'^2+k_1^2+2m_\pi^2) \frac{1}{2\ell +1} [\ell Q_{W0}^{\ell-1}(k_1',k_1)+(\ell+1) Q_{W0}^{\ell+1}(k_1',k_1)] \nonumber\\
 & & -k_1'k_1\frac{1}{2\ell+1} \left\{\frac{(\ell-1)\ell}{2\ell-1} Q_{W0}^{\ell-2}(k_1',k_1)
 +\left(\frac{(\ell+1)^2}{2\ell+3} +\frac{\ell^2}{2\ell-1}\right) Q_{W0}^{\ell}(k_1',k_1)
 +\frac{(\ell+2)(\ell+1)}{2\ell+3} Q_{W0}^{\ell+2}(k_1',k_1) \right\}.
\end{eqnarray}
The tensor components of the $c_4$ interaction of $V_C$, Eq. (A10), are
\begin{eqnarray}
 & & 2\frac{c_4 g_A^2}{4f_\pi^4} \frac{2}{3} (S_{12})_{\ell 1J}^{\ell'}(\bftau_1\cdot\bftau_2)
 \left[ \frac{1}{2k_1'k_1} (k_1'^2Q_{\ell}(z)+k_1^2Q_{\ell'}(z)-2k_1'k_1 Q_J(z))
\left( \frac{1}{2}\rho_0-m_{\pi}^2(F_0(K_1')+F_0(k_1)) \right) \right.\nonumber\\
 & & +\frac{k_1'k_1}{2J+1} (Q_{W0}^{J+1}(k_1',k_1)-Q_{W0}^{J-1}(k_1',k_1)) \nonumber \\
 & & +(F_0(k_1')-2F_1(k_1')) \left\{\frac{k_1'^2}{4k_1'k_1}(-k_1'^2+k_1^2+m_{\pi}^2)Q_{\ell}(z)
 -\frac{1}{4}(-k_1'^2+k_1^2+m_{\pi}^2)Q_J(z)-k_1'^2\delta_{\ell 0}+\frac{1}{2}k_1'k_1\delta_{J0}\right\} \nonumber \\
 & & +(F_0(k_1)-2F_1(k_1)) \left\{\frac{k_1^2}{4k_1'k_1}(k_1'^2-k_1^2+m_{\pi}^2)Q_{\ell'}(z)
 -\frac{1}{4}(k_1'^2-k_1^2+m_{\pi}^2)Q_J(z)-k_1^2\delta_{\ell 0}+\frac{1}{2}k_1'k_1\delta_{J0}\right\} \nonumber \\
 & & +\frac{1}{3}F_3(k_1') \left\{-\frac{3}{2}k_1'^2\delta_{\ell 0} +\frac{3}{2} k_1'k_1\delta_{J0}
  +\frac{k_1'^2}{4k_1'k_1}(-k_1'^2+3k_1^2+3m_{\pi}^2)Q_{\ell}(z)-\frac{1}{4}(k_1'^2+3k_1^2+3m_{\pi}^2)Q_J(z)
+\frac{1}{2}k_1'k_1Q_{\ell'}(z) \right\}\nonumber \\
 & & +\frac{1}{3}F_3(k_1) \left\{-\frac{3}{2}k_1^2\delta_{\ell 0} +\frac{3}{2} k_1'k_1\delta_{J0}
  +\frac{k_1^2}{4k_1'k_1}(-k_1^2+3k_1'^2+3m_{\pi}^2)Q_{\ell'}(z)-\frac{1}{4}(k_1^2+3k_1'^2+3m_{\pi}^2)Q_J(z)
+\frac{1}{2}k_1'k_1Q_{\ell}(z) \right\}\nonumber \\
 &&\left. -\frac{1}{3}F_2(k_1') \frac{k_1'^4}{2k_1'k_1}Q_{\ell}(z)-\frac{1}{3}F_2(k_1) \frac{k_1^4}{2k_1'k_1}Q_{\ell'}(z) \right]
\end{eqnarray}
for $\ell'=\ell\pm 1$ ($J=\ell\pm 1$) and
\begin{eqnarray}
 & &  2\frac{c_4 g_A^2}{4f_\pi^4} \frac{2}{3} (S_{12})_{\ell 1J}^{\ell'}(\bftau_1\cdot\bftau_2)
 \left[ \left(\frac{1}{2}\rho_0-m_{\pi}^2(F_0(k_1')+F_0(k_1))\right)\left\{\frac{k_1^2+k_1^2}{2k_1'k_1}Q_{\ell}(z)
  -\frac{1}{2}\frac{2\ell+3}{2\ell+1}Q_{\ell-1}(z)-\frac{1}{2}\frac{2\ell-1}{2\ell+1}Q_{\ell+1}(z)\right\} \right. \nonumber \\
  & & +k_1'k_1\left\{ \left(\frac{(2\ell+1)^2}{(2\ell-1)(2\ell+3)}-2\right)Q_{W0}^{\ell}(k_1',k_1)
 +\frac{(2\ell+3)(\ell-1)}{(2\ell+1)(2\ell-1)}Q_{W0}^{\ell-2}(k_1',k_1)
 +\frac{(2\ell-1)(\ell+1)}{(2\ell+1)(2\ell+3)}Q_{W0}^{\ell+2}(k_1',k_1)\right\} \nonumber \\
 & & +(F_0(k_1')-2F_1(k_1'))\left\{ -\frac{1}{2}k_1'^2\delta_{\ell 0} +\frac{5}{12}k_1'k_1\delta_{\ell 1}
+\frac{k_1'^2}{4k_1'k_1}(-k_1'^2+k_1^2+m_{\pi}^2)Q_{\ell}(z) \right. \nonumber\\
 & & \left. -\frac{1}{8(2\ell+1)}(-k_1'^2+k_1^2+m_{\pi}^2)((2\ell+3)Q_{\ell-1}(z)+(2\ell-1)Q_{\ell+1}(z)) \right\}
 +\frac{1}{3}F_3(k_1') \left\{ -\frac{3}{2}k_1'^2\delta_{\ell 0} +\frac{5}{4}k_1'k_1\delta_{\ell 1} \right. \nonumber \\
 & &  \left. +\frac{k_1'^2}{4k_1'k_1}(-k_1'^2+5k_1^2+3m_{\pi}^2)Q_{\ell}(z))
-\frac{1}{8(2\ell+1)}(k_1'^2+3k_1^2+3m_{\pi}^2)((2\ell+3)Q_{\ell-1}(z)+(2\ell-1)Q_{\ell+1}(z)) \right\} \nonumber\\
 & & -\frac{1}{3}F_2(k_1')\frac{k_1'^4}{2k_1'k_1}Q_{\ell}(z)
 +(F_0(k_1)-2F_1(k_1))\left\{ -\frac{1}{2}k_1^2\delta_{\ell 0} +\frac{5}{12}k_1'k_1\delta_{\ell 1}
+\frac{k_1^2}{4k_1'k_1}(k_1'^2-k_1^2+m_{\pi}^2)Q_{\ell}(z) \right. \nonumber\\
 & & \left. -\frac{1}{8(2\ell+1)}(k_1'^2-k_1^2+m_{\pi}^2)((2\ell+3)Q_{\ell-1}(z)+(2\ell-1)Q_{\ell+1}(z)) \right\}
 +\frac{1}{3}F_3(k_1) \left\{ -\frac{3}{2}k_1^2\delta_{\ell 0} +\frac{5}{4}k_1'k_1\delta_{\ell 1} \right. \nonumber \\
 & &  \left. +\frac{k_1^2}{4k_1'k_1}(-k_1^2+5k_1'^2+3m_{\pi}^2)Q_{\ell}(z))
-\frac{1}{8(2\ell+1)}(k_1^2+3k_1'^2+3m_{\pi}^2)((2\ell+3)Q_{\ell-1}(z)+(2\ell-1)Q_{\ell+1}(z)) \right\} \nonumber\\
 & & -\frac{1}{3}F_2(k_1)\frac{k_1^4}{2k_1'k_1}Q_{\ell}(z)
\end{eqnarray}
 for $\ell'=\ell=J\pm 1$. There is no spin-orbit component from the $c_4$ interaction of $V_C$, Eq. (A10).

The central component of the $V_D$ interaction, Eq. (A11), is 
\begin{eqnarray}
 && \frac{g_A}{8f_\pi^2} \frac{c_D}{f_\pi^2 \Lambda_\chi}
  \frac{1}{3}(\bfsigma_1\cdot \bfsigma_2)(\bftau_1\cdot \bftau_2)\left\{\frac{\rho_0m_\pi^2}{2k_1'k_1} Q_\ell (x)
  -\delta_{\ell 0} 2m_\pi^2( F_0(k_1)+ F_0(k_1'))\right\}\nonumber\\
  & &+ 3\frac{g_A}{8f_\pi^2} \frac{c_D}{f_\pi^2 \Lambda_\chi} \delta_{\ell 0} 
  \{\rho_0-2m_\pi^2(F_0(k_1)+F_0(k_1'))\}.
\end{eqnarray}
The tensor component for the initial $\ell$ and the final $\ell'=\ell\pm 1$ ($J=\ell\pm 1$) becomes
\begin{eqnarray}
 & & 2\frac{g_A}{8f_\pi^2} \frac{c_D}{f_\pi^2 \Lambda_\chi} (\bftau_1\cdot \bftau_2)\frac{1}{3} S_{12}
 \{ F_0(k_1)-2F_1(k_1)+F_3(k_1) +F_0(k_1')-2F_1(k_1')+F_3(k_1')\}
 (k_1'^2 \delta_{\ell 0}+k_1^2 \delta_{\ell' 0} -2k_1'k_1 \delta_{J 0})\nonumber \\
 & & -\frac{g_A}{8f_\pi^2} \frac{c_D\rho_0}{f_\pi^2 \Lambda_\chi}  (\bftau_1\cdot \bftau_2)
 \frac{1}{3} S_{12} \left\{\frac{k_1'^2}{2k_1'k_1} Q_\ell (x)+\frac{k_1^2}{2k_1'k_1} Q_{\ell'} (x)
 - Q_J (x) \right\},
\end{eqnarray}
and for $\ell'=\ell=J, J\pm 1$
\begin{eqnarray}
 & & 2\frac{g_A}{8f_\pi^2} \frac{c_D}{f_\pi^2 \Lambda_\chi} (\bftau_1\cdot \bftau_2)\frac{1}{3} (S_{12})_{\ell 1J}^{\ell'}
 \{ F_0(k_1)-2F_1(k_1)+F_3(k_1) +F_0(k_1')-2F_1(k_1')+F_3(k_1')\} \nonumber \\
 & & \hspace*{10cm} \times \left\{ k_1'^2 \delta_{\ell 0}+k_1^2 \delta_{\ell' 0}
     -\frac{5}{3}k_1'k_1 \delta_{\ell 1}\right\} \nonumber \\
 & & -\frac{g_A}{8f_\pi^2} \frac{c_D\rho_0}{f_\pi^2 \Lambda_\chi}  (\bftau_1\cdot \bftau_2)
 \frac{1}{3} S_{12}
 \left\{\frac{k_1'^2}{2k_1'k_1} Q_\ell (x)+\frac{k_1^2}{2k_1'k_1} Q_{\ell'} (x)
 -\frac{1}{2}\left(\frac{2\ell+3}{2\ell+1}Q_{\ell-1}(z)+\frac{2\ell-1}{2\ell+1} Q_{\ell+1}(z)\right) \right\}.
\end{eqnarray}
There is no spin-orbit component from the $c_4$ interaction of $V_D$, Eq. (A11).

Finally, the $V_E$ interaction, Eq. (A12), gives only an $\ell=0$ central component; namely,
$-6\frac{C_E\frac{1}{4}\rho_0}{f_\pi^4\Lambda_\chi}$ both for $^1S_0$ and $^3S_1$ channels.
\bigskip
\end{widetext}


\begin{thebibliography}{99}
\bibitem{BLM54} K.A. Brueckner, C.A. Levinson, and H.M. Mahmoud, Phys. Rev. {\bf 95}, 217
(1954).
\bibitem{DAY67} B.D. Day, Rev. Mod. Phys. {\bf 39}, 719 (1967).
\bibitem{BET71} H.A. Bethe, Ann. Rev. Nucl. Sci. {\bf 21}, 93 (1971). 
\bibitem{PW79} V.R. Pandharipande and R.B. Wiringa, Rev. Mod. Phys. {\bf 51}, 821 (1979).
\bibitem{BB12} M. Baldo and G.F. Burgio, Rep. Prog. Phys. {\bf 75}, 026301 (2012).
\bibitem{BM90} R. Brockmann and R. Machleidt, Phys. Rev. C {\bf 42}, 1965 (1990).
\bibitem{MACH} R. Machleidt and D.R. Entem, Phys. Rep. {\bf 503}, 1 (2011).
\bibitem{EGM05} E. Epelbaum, W. G\"{o}ckle, and U.-G. Mei{\ss}ner,
Nucl. Phys. A {\bf 747}, 362 (2005).
\bibitem{EHM09} E. Epelbaum, H.-W. Hammer, and U.-G. Mei{\ss}ner,
Rev. Mod. Phys. {\bf 81}, 1773 (2009).
\bibitem{KL53} A. Klein, Phys. Rev. {\bf 90}, 1101 (1953).
\bibitem{FM57} J. Fujita and H. Miyazawa, Prog. Theor. Phys {\bf 17}, 366 (1957).
\bibitem{WIR00} R.B. Wiringa, S.C. Pieper, J. Carlson, and V.R. Pandharipande,
Phys. Rev. C {\bf 62}, 014001 (2000).
\bibitem{KNE12} N. Kalantar-Nayestanaki, E. Epelbaum, J.G. Messchendorp,
and A. Nogga, Rep. Prog. Phys. {\bf 75}, 016301 (2012).
\bibitem{KAT74} T. Kasahara, Y. Akaishi, and H. Tanaka, Prog. Theor. Phys. Suppl. {\bf 56}, 96 (1974).
\bibitem{FP81} B. Friedman and V.R. Pandharipande, Nucl. Phys. A {\bf 361}, 361 (1981).
\bibitem{APR98} A. Akmal, V.R. Pandharipande, and D.G. Ravenhall, Phys. Rev. C {\bf 58}, 1804 (1998).
\bibitem{BSFN05} S.K. Bogner, A. Schwenk, R.J. Furnstahl, and A. Nogga,
Nucl. Phys. A {\bf 763}, 59 (2005).
\bibitem{HEB11} K. Hebeler, S.K. Bogner, R.J. Furnstahl, A. Nogga,
and A. Schwenk, Phys. Rev. C {\bf 83}, 031301(R) (2011).
\bibitem{HS10} K. Hebeler and A. Schwenk, Phys. Rev. C {\bf 82}, 014314 (2010).
\bibitem{TKHS13} I. Tews, T. Kr\"{u}ger, K. Hebeler, and A. Schwenk, Phys. Rev. Lett. {\bf 110},
032504 (2013).
\bibitem{CORA13} L. Coraggio, J.W. Holt, N. Itaco, R. Machleidt, and F. Sammarruca, nucl-th/1209.5537.
\bibitem{MK12} M. Kohno, Phys. Rev. C {\bf 86}, 061301(R) (2012).
\bibitem{SCC12} F. Sammarruca, B. Chen, L. Coraggio, N. Itaco, and R. Machleidt,
Phys. Rev. C {\bf 86}, 054317 (2012).
\bibitem{LNR71} B.A. Loiseau, Y. Nogami, and C.K. Ross, Nucl. Phys. A {\bf 165}, 601 (1971);
Erratum A {\bf 176}, 665 (1971).
\bibitem{HKW10} J.W. Holt, N. Kaiser, and W. Weise, Phys. Rev. C {\bf 81}, 024002 (2010).
\bibitem{KO12} M. Kohno and R. Okamoto, Phys. Rev. C {\bf 86}, 014317 (2012).
\bibitem{HT70} M.I. Haftel and F. Tabakin, Nucl. Phys. A {\bf 158}, 1 (1970).
\bibitem{SOKN00} K. Suzuki, R. Okamoto, M. Kohno, and S. Nagata,
Nucl. Phys. A {\bf 665}, 92 (2000).
\bibitem{YF00} Y. Fujiwara, M. Kohno, T. Fujita, C. Nakamoto, and Y. Suzuki,
Prog. Theor. Phys. {\bf 103}, 755 (2000).
\bibitem{AV18} R.B. Wiringa, V.G.J.Stoks, and R. Schiavilla,
Phys. Rev. C {\bf 51}, 38 (1995).
\bibitem{NSC} T. A. Rijken, V. G. J. Stoks, and Y. Yamamoto, Phys. Rev. C {\bf 59}, 21 (1999).
\bibitem{CDB} R. Machleidt, Phys. Rev. C {\bf 63}, 024001 (2001).
\bibitem{GP77} D. Gogny and R. Padjen, Nucl. Phys. A {\bf 293}, 365 (1977).
\bibitem{S13} T. Sasaki, N. Yasutake, M. Kohno, H. Kouno, and M. Yahiro, arXiv:1307.0681.
\bibitem{BM} A. Bohr and B. Mottelson, Nuclear structure, viol. I (Benjamin, Reading, Mass.,1969).
\bibitem{YRS} Y.R. Shimizu, private communication.
\end{thebibliography}
\end{document}